# SR-FTIR MICROSCOPY AND FTIR IMAGING IN THE EARTH SCIENCES

Della Ventura, G[1,2*], Marcelli, A[2], Bellatreccia, F[1,2]

1) Dipartimento Scienze, Università di Roma Tre, Largo S. Leonardo Murialdo 1, I-00146 Roma (Italy)
giancarlo.dellaventura@uniroma3.it, fabio.bellatreccia@uniroma3.it
2) INFN - Laboratori Nazionali di Frascati, Via E. Fermi 40, I-00044 Frascati (Rome), Italy
marcelli@lnf.infn.it

*Corresponding author:
Tel: +39-06-57338020
Fax: +39-06-57338201
e-mail: dellaven@uniroma3.it

# INTRODUCTION

Infrared spectroscopy was developed at the beginning of the XX century for analytical chemical purposes and it is remarkable that the first contribution of the first issue of Physical Review, one of the first and among the most important physical magazines, published in 1883, was devoted to "a study of the transmission spectra of certain substances in the infrared" that included also a plate of a quartz rock crystal (Nichols 1883). In 1905 William W. Coblentz released the very first database of IR spectra where the characteristic wavelength at which various materials absorbed the IR radiation were listed. For more than a century IR spectroscopy has been considered as a powerful analytical tool for phase identification and to characterize the structural features and quantify molecules or molecular arrangements in solids, and also in liquids and gases. This technique has been used extensively by organic chemists, and since the 1950s it has been recognized as a fundamental technique in mineralogical and Earth sciences in conjunction with X-ray diffraction (Keller and Pickett 1949, 1950, Launer 1952, Adler and Kerr 1965). The first "encyclopedia" of IR spectra of minerals appeared in 1974 (Farmer 1974) and is still a primary reference for those using infrared spectroscopy as a tool in material science.

In the late '60s – beginning of '70s a new IR technique, Fourier-transform infrared (FTIR) spectroscopy, was developed and later a new class of spectrometers was commercially available, based upon the combination of a FFT (Fast-Fourier-Transform) algorithm and computers. The advantage of these instruments over the previously used wavelength-dispersive spectrometers is fully described in several books, e.g. Griffiths and de Haseth (1986) and Smith (1996). Briefly, in FTIR spectrometers IR light is focused through an interferometer and then through the sample. A moving mirror inside the apparatus alters the distribution of infrared light (wavelength) that passes through the interferometer. The recorded signal, called an interferogram, represents light output as a function of mirror position, which correlates with wavelength. A FFT data-processing transforms the raw data into the sample spectrum. A major advantage of the FTIR spectrometers is that the information in the *entire* frequency range is collected simultaneously, improving both speed and signal-to-noise ratio (SNR).

During the last decades, several books have been devoted to the application of spectroscopic methods in mineralogy, e.g. Volume 18 of Reviews in Mineralogy (Hawthorne 1988). Several short courses (e.g. Beran and Libowitzky 2004) and meetings have addressed particular aspects of spectroscopy, such as the analysis of hydrous components in minerals and Earth materials (e.g. Keppler and Smyth 2006). In these books, complete treatment of the infrared theory and practical aspects of instrumentation and methods, along with an exhaustive list of references, can be found.

The present chapter is intended to cover those aspects of infrared spectroscopy that have been developed in the past decade and are not included in earlier reviews such as Volume 18 of Reviews in Mineralogy. These new topics involve primarily: (1) the use of synchrotron radiation (SR), which, although not a routine method, is now rather extensively applied in infrared studies, in particular those requiring ultimate spatial and time resolution and the analysis of extremely small samples (a few tens of micrometers); (2) the development of imaging techniques also for foreseen time resolved studies of geo-mineralogical processes and environmental studies.

There are now several synchrotron beamlines around the world that are dedicated to IR spectroscopy and microscopy, hosting an ever-increasing number of multidisciplinary users. Interest in synchrotron-IR radiation (SR-IR) increased in the last 20 years because of its unique properties and effective advantages, including the possibility to perform experiments using an intense fully-linear polarized light source.

A non-thermal SR source is an intense source whose emission at long wavelengths is asymptotic and, for accumulated electrons at E > 0.5 GeV, does not depend on the energy. Its intensity is also proportional to the large horizontal opening angle and to the current circulating in the storage ring. The unique features of SR-IR radiation overcome limitations of conventional "benchtop" instruments and open new fields of application, in particular those associated to extreme experimental conditions (e.g., high-pressure, cryogenic/high temperatures).

As we will discuss in more details later, the main advantage of synchrotron radiation over conventional black body sources is its brilliance, which is defined as the photon flux density normalized to the source area and to the horizontal and vertical angle of emission (photons/mm$^2$/mrad$^2$). The gain in brilliance of an infrared SR source with respect to a conventional source is from 100-1000 times going from the near-IR (NIR) to the far-IR (FIR), and increases when working at small apertures. It is worth stressing, however, that the best advantage of a SR source over a conventional source is obtained when using a confocal microscope or when performing time-resolved experiments. A microscope has a confocal geometry when the objective and the collector share the same focus at the sample location and both have a small aperture, placed at the conjugated focus, limiting the illuminated or the detected area of the sample, respectively. The use of small apertures such as pinholes allows images with a high spatial resolution to be obtained, but at the cost of a low throughput. In this latter case the signal-to-noise ratio (SNR) must ensure the recognition of small features in the image. The drawback of the use of pinholes or extremely small rectangular apertures (say < 20 $\mu m^2$) in a microscope is the dramatic degradation of the SNR and the need to increase the collection time unless using brilliant SR sources.

In addition, the recent advent of area-detectors ("focal-plane-array, FPA") has revolutionized the world of FTIR spectroscopy. These arrays are composed of small IR detectors (pixels) a few tens of microns or less in dimension, and allow the acquisition of thousands of IR spectra simultaneously generating mid-IR (MIR) images with a high resolving power. The optical system of an IR microscope equipped with a FPA is an *apertureless* imaging system, whose ultimate spatial resolution is comparable but never better then that attainable by confocal microscopes. Nevertheless, FPAs enable different imaging modes at the resolution of a few microns and, thanks to their sensitivity and read-out speed, massive and fast data collection within minutes are possible (Bhargava and Levin 2001, Petibois and Déléris 2006, Petibois et al. 2010a).

For Earth science materials, SR-IR has been applied in studies where an improved spatial resolution was needed, for example for fine-grained synthetic crystals, or when high brilliance was needed, such as in high-pressure (HP) studies using diamond-anvil cells (DAC) (Scott et al. 2007, Noguchi et al. 2012). The new imaging capabilities of FTIR detectors were used to address features such as zoning of volatile species across the sample or possible configurational changes of structurally-bound carbon molecular species (e.g., $CO_2$ *vs* $CO_3$) during the crystal growth. Such features, which are barely accessible with micro-analytical techniques, may provide constraint in terms of physico-chemical parameters relating to the conditions of formation of the samples, and important information on the evolution of geological systems *vs.* time. Some significant recent studies will be reviewed here that explore the use of FTIR imaging to investigate experimental systems and to monitor processes in real time.

## FTIR MICROSCOPY AND IMAGING TECHNIQUES

More than a century ago, Ernst Abbe showed experimentally that the resolving power of an optical instrument is subject to a physical limit, i.e., the diffraction limit, which cannot be overcome by designing better objective lenses. The diffraction limit is ~ $2\lambda/NA$ where *NA* is the numerical aperture of the microscope objective. Working with commercial objectives having *NA* ~0.6, the resolution is 3 to 4 times the wavelength (~ 1.7 μm at 4000 cm$^{-1}$). In a confocal microscope, where the objectives are placed both before and after the sample, the spatial resolution can be improved to ~ $\lambda/2$ (Minsky 1988).

The goal of FTIR microscopy is to achieve the best spatial resolution at the sample location, i.e., to allow imaging to the diffraction limit. However, as it has been shown recently (Levin and Bhargava 2005, Levenson et al. 2006, Bhargava and Levin 2007), the concept of spatial resolution in IR imaging is not an independent parameter: it is diffraction limited and is also affected by both the optical design (e.g., objectives and apertures) and contrast. To avoid chromatic aberration, IR microscopes are equipped with two Schwarzschild objectives, which are based on two spherical mirrors centred on the same optical axis and allow magnifications of 15x or 36x with numerical apertures (*NA*) in the range 0.3-0.7. The magnification, the parameter that determines how much small features in the specimen can be enlarged, must be adjusted in proportion to distance (see Fig. 1). To increase the magnification, the objective of the microscope has to be positioned nearer to the specimen to be observed. The correlation between the magnification and the *NA* of the objective and the condenser of a microscope is: M= (f'/f)= (*NA*/*NA*'), where f and f' are the focal distance of the objective and the condenser microscope optical elements, respectively. This definition is practical and useful because it is independent of any specific characteristics of the optical system (M $\propto$ *NA*). The concept of the lateral spatial resolution of an image, i.e. the spatial resolution in the plane perpendicular to the propagation of the light through the specimen, has ~~also~~ to be related to a resolution criterion, i.e., how we may clearly separate two closely spaced objects inside a sample. According to the Rayleigh criterion, two points can be separated when the central maximum of the first Airy disk is placed at a distance greater than the radius of the first minimum of the Airy disk (see Figure 2). When a point source of monochromatic radiation goes through a microscope, the Airy pattern is observed at the beam focus. The central circular area, i.e., the Airy disc, of the point spread function (PSF) in Figure 2 is characterized by a radius r = 0.61($\lambda$/*NA*), where $\lambda$ is the wavelength and *NA* is the numerical aperture of the microscope. However, the Rayleigh criterion is valid only when the two points to be separated both sit on an identical and negligible background and have the same intensity, a condition that corresponds to a minimum contrast of 26.4% (Levenson et al. 2006, Pawley 2006).

If we consider a standard optical microscope illuminating a large sample region, the ultimate resolution is achieved by the ''detection system'' made by the objective optics and our eye. Similarly, in an IR microscope, when using an area-detector, i.e., a focal-plane array (FPA) detector, the maximum spatial resolution is determined by the magnification of the optical system, but the size of the individual pixel of the detector is also a relevant parameter (Miller and Smith 2005). FPA detectors are still not widespread because almost all IR microscopes are equipped with single-element MCT detectors.

In a microscope, reducing with a field stop the field of view (FOV) at the sample location, for example by an aperture placed at an intermediate focal point, the spatial resolution is determined by the fraction of light from each point of the specimen that reaches the detector, i.e., its sensitivity pattern. As outlined above, if we reduce the illumination region while maintaining a homogeneous illumination, then the spatial resolution is a function of the number of photons that illuminate the small selected region of the specimen. To probe this small region of a specimen, the microscope must illuminate it and simultaneously detect the light. A confocal microscope has apertures for both the illumination and detection systems. It achieves the ultimate spatial resolution by reducing to the same size both the illuminated region and the region to be detected. An image of the specimen under analysis can be then collected, although with much longer times, by rastering the sample through the focus of the confocal microscope. In this case, because of their intrinsic noise, large single-element detectors (50 x 50 μm or larger) are not suitable to work at high spatial resolution and the dimension of the single-element detector is the real bottleneck.

In contrast to confocal microscopes with single-element detectors, combining a SR source with a two-dimensional FPA detector is an efficient way to take advantage of a SR source for imaging and time resolved experiments. Using area-detectors we may collect FTIR spectra simultaneously on large areas (several tens of $\mu m^2$) depending on the magnification. Images containing hundreds of points are obtained within minutes compared with longer acquisitions (hours) typically required by a single-element detector mapping. Although, in principle, a thermal source may achieve a sub-second resolution for a single spectrum, using the raster scan method images can be collected at a much slower time. Moreover, working with a high current synchrotron radiation facility, the intense IR beam can be shaped to illuminate only a limited number of pixels of a FPA detector increasing the SNR ratio of the image achieved with high contrast (Petibois et al., 2010a). For these reasons, FPAs are used extensively in biological studies (e.g., Petibois et al. 2009), while very few studies of Earth materials have been made so far, despite its wide potential applications.

When considering resolution, great emphasis is given to the lateral resolution. However, to interpret an image it is also important to consider the axial resolution, i.e., the axial resolving power of the objective measured along the optical axis. As with the lateral resolution, the axial resolution (Fig. 2) also depends on the numerical aperture (*NA*) of the objective (Born and Wolf 1997, Inoué 1995). Lateral and axial spatial resolution for a confocal microscope are reduced by ~ 30% if compared with a corresponding microscope working with a wide-field illumination (Fig. 2). A large *NA* improves both the lateral and axial spatial resolution of the microscope although the *NA* of the microscope objective is much more effective for the axial resolution ($NA^2$). Knowledge of the axial resolution is fundamental to performing "optical sectioning", a technique that makes possible a three-dimensional reconstruction of a sample *via* non-destructive depth profiling, collecting images as the focus is moved deeper into the sample. This method offers clear advantages with respect to mechanically cutting a cross-section (microtomy), avoiding in particular the need for any destructive sample preparation.

In this context, it is worth also mentioning the opportunity offered by scanning near-field optical microscopy (SNOM), a technique which allows optical imaging that can be extended to the FTIR domain with a resolution down to few tenths of nm, i.e. well beyond the Abbe diffraction limit (e.g., Cricenti et al. 2002, Vobornik et al. 2005). SNOM systems are based on optical waveguides with a nano-aperture much smaller than the light wavelength λ. The use of a nano-aperture as a local probe makes it possible to overcome the λ/2 diffraction limit that occurs if the light is detected in far field, i.e., when the detector distance is much larger than the wavelength λ of the light used in the experiment. However, to the authors' knowledge, no application of this technique to geological samples has been published so far, except some work aimed at studying diffuse nanoparticles in meteorite's metal inclusions thought to be responsible for the problem of asteroid reddening (Pompeo et al. 2010). Imaging can be obtained with an *apertureless* scanning near-field optical microscope, leading to an extreme spatial resolution at IR wavelengths. However, we need to point out that it is practically impossible to map a large sample with SNOM optics. Moreover, because optical methods measure the index of refraction of a material, the real part returns only the sample topography while the imaginary part is proportional to the sample absorption, and so when looking at extremely small areas the local absorption is weak. In other words, if the signal is too small to be detected or comparable to the noise, differential methods such as photoacoustic or photothermal detection are more efficient. A new IR spectromicroscopy technique, based on the coupling between a tunable free-electron IR laser and an Atomic Force Microscope in the IR range (AFMIR) was recently setup at the CLIO free-electron laser facility (Dazzi et al., 2005, 2007a,b; Ortega et al., 2006) allowing IR mapping at the nm scale. Detection is performed directly by an AFM tip in the contact mode, probing the local thermal expansion of the sample, irradiated at the wavelength of specific absorption bands. As the duration of expansion and relaxation of the sample is always shorter than the response time of the cantilever in contact, by recording the amplitude of the cantilever oscillations it is possible to measure the corresponding IR absorption as a function either of space or wavelength. A spatial resolution around 50 nm has been achieved in the mid-IR region at λ=22 μm (Houel et al., 2009), although, because of the inherent near-field character, the sensitivity of this technique is reduced for objects located below the surface.

For FTIR imaging in transmission mode, the sample preparation is ~~another~~ crucial aspect to consider. The specimen must be prepared as a slice with parallel and doubly-polished surfaces to allow the IR beam to pass through the sample without scattering. A fragment of each material under investigation, either a rock or a single-crystal, is cut into a block and polished on both sides using different kinds of abrasives, e.g. silicon carbide or alumina powders, diamond pastes, or abrasives plastic foils. Final polishing is obtained with a 1.0 to 0.25 μm grit. During the grinding, the sample is usually mounted on a glass slide with easily removable glues, such as Crystalbond™. The final thickness depends on the type of sample and on the problem under investigation. At the end, the slice is removed from the glass and carefully washed from any residual from epoxy or binding agent. The result is a thin (typically in the range 300–15 μm) freestanding doubly polished wafer. For quantitative purposes, an additional critical point is the measurement of the slice thickness. This is usually achieved using a micrometer, and checked with the IR microscope using standard slabs with known thickness as reference. When high precision is required, measurements using a scanning electron microscope (SEM) or an optical profilometer (e.g. Della Ventura et al. 2012) are preferred. Non-conventional techniques recently available for sample preparation involve the use of focused ion beam (FIB) instruments to cut extremely thin slices from target sample locations (Koch-Müller et al. 2004, see also Wirth 2004 and Marquardt and Marquardt 2012). An example of FIB machining to prepare oriented very thin crystal sections for polarized-light FTIR measurements is given in Figure 3.

# SYNCHROTRON-RADIATION FTIR SPECTROSCOPY IN MINERAL SCIENCES

## Introduction

As a light source for spectroscopy, synchrotron radiation has several key properties: (1) high brilliance, (2) a continuous distribution of the intensity over the entire spectral range, (3) a pulsed nature, (4) a high degree of polarization and (5) a high stability. Synchrotrons and storage rings use magnetic structures, i.e., dipoles, quadrupoles and sextupoles to bent and focus electrons to circulate along closed orbits. Radiation is emitted when particles travelling at relativistic velocities are deflected by a constant magnetic field as it occurs inside a bending magnet. This is the classical synchrotron radiation bending magnet emission whose spectral range is characterized by a broad spectrum extending from the microwave to the X-ray domain (Hofmann 2004). Radiation is also emitted by electron travelling through a fringing field of a dipole magnet. This emission is called "Edge Radiation" (ER). For electrons at a relativistic speed, the fringe field is like a "sharp edge transition" from a zero field to a full field value and the associated impulsive acceleration originates the emission of light. The emission spectrum of ER is limited and does not extend to the X-ray domain; however, in the low-frequency (IR, THz) range it is comparable to the standard synchrotron radiation spectrum (Geloni et al. 2009a, 2009b).

The main advantage of synchrotron radiation is its brilliance. In spite of it, large angles are required to extract from bending magnets long wavelength radiation such as IR radiation, because the "natural" opening angle increases up to several tens milliradians in the far-IR range. On the contrary, long-wavelength radiation emitted as Edge Radiation is characterized by a significantly smaller opening angle than standard bending-magnet radiation. Long wavelength SR sources may have a strong potential for IR spectroscopy or imaging techniques (Petibois et al. 2009) particularly considering that both ER and bending magnet radiation have equivalent brilliance. The history of SR utilization in the long wavelength region (from micrometer to millimeter waves) is more recent than that in the short wavelength domain (Marcelli and Cinque 2011). In fact, SR sources are some order of magnitude more brilliant than a conventional black body source in the same spectral range as shown in Figure 4. A highly brilliant source is not a priority for a spectroscopic measurement, while becomes mandatory in microscopy where a spot size down to the diffraction limit is desired. However, the SNR decreases drastically as apertures are reduced to confine the beam into a small area to increase the lateral resolution of an image, as shown in Figure 5. The use of IR microscopes coupled to SR sources may guarantee outstanding results just because the high brilliance of this source allows reducing the microscope apertures down to the *diffraction limit*, i.e., down to a few microns in the mid-IR region, still assuring a good SNR (Carr 2001).

Because the electrons in the storage ring do not form a continuous distribution around the orbit but are stored and travel as bunches in a circular accelerator, the resulting emission has a time structure. The SR is indeed emitted as light pulses characteristic of each accelerator, being another interesting feature of this source for time resolved experiments. The bunch length ranges from tens of ps to a few ns, values hardly to be obtained with conventional sources (Mills 1984, Xu et al. 2011, Innocenzi et al. 2009).

An additional important feature of synchrotron radiation is its ~~strong~~ polarization. Owing to its relativistic character, the SR emission is linearly polarized in the orbital plane. Since the early pioneering work in the visible region, studies performed both in the VUV and X-ray regions demonstrated that when observed at a particular angle over the orbit, the radiation is circularly polarized. The polarization properties of the SR at long wavelengths are very promising for many applications, in particular in the far-IR region. By placing a slit on the exit port, one can in principle select the desired degree of circular polarization and the flux of the emitted radiation. Dealing with polarization, we may define three polarization rates, of which two are more important: the linear polarization (P1) and the circular polarization (P3). Experimental observations indicate that values up 80% of circularly polarized light can be obtained using a slit that selects ~50% of the total flux available (Cestelli Guidi et al. 2005).

In SR-IR beamlines, commercial optical benches and microscopes are installed and the synchrotron beam replaces the conventional black body source. The beam is focused in the upper aperture of the microscope (Fig. 6) and is projected onto the sample plane by a Catoptric objective; the size of the aperture determines the lateral resolution that one can attain in the analysis; the problems associated with the SNR and the resolution in these conditions have been introduced before. More details can be found in Marcelli and Cinque (2011).

**Applications in mineral sciences**

For materials of interest in the Earth Sciences SR-FTIR has been mostly applied in the study of inclusions within minerals, in the analysis of $H_2O$/OH in nominally anhydrous minerals (NAMs), of H and C molecules of interplanetary dust particles (IDPs) and in high-pressure studies. Few applications in the field of cultural heritage have also been reported.

Guilhamou et al. (1998) and Bantignies et al. (1998) were among the very first to explore the potentiality of SR-FTIR microspectrometry for the study of samples of geological interest (Fig. 7). In particular, Guilhamou et al. (1998) addressed the analysis of hydrocarbon-rich fluid inclusions in siliceous diagenetic materials, aimed at constraining petroleum formation and migration. Using a beam size of 3 x 3 $\mu m^2$ they were able to obtain high SNR spectra for an accurate chemical identification of aliphatic components, $CO_2$ and water entrapped in fluorite cementing the host rock. In the same paper they also reported preliminary tests for the analysis of $CO_2$ and $H_2O$ in glass inclusions within olivines from tholeitic basalts from South Vietnam and the detection of trace OH in NAMs.

Guilhaumou et al. (2005) studied melt inclusions within garnets and exsolved pyroxenes from deep-seated ultramafic xenoliths uplifted in the Jagersfontein kimberlite pipe (South Africa). Petrographical and mineralogical data showed these samples to record evidence of multistage fluid-rock interactions during their ascent from the asthenosphere through the lithosphere. Primary and secondary OH-bearing melt inclusions were recognized, the first representing the result from an early partial melting of the garnet host, and the second originating from volume change of the hydrated pyroxene during ascent. The data were interpreted as evidence for unusual water contents in the primary ultradeep-seated garnets. Evidence of a carbonaceous magma rich in dissolved $CO_2$ was provided by the presence of secondary inclusions containing a $CO_3$-bearing glassy phase associated with a vapor phase rich in $CO_2$ and $H_2O$.

Koch-Müller et al. (2004) and Koch-Müller et al. (2006) studied, using SR-FTIR, the hydroxyl content of pyroxenes and olivines from high-pressure occurrences. Three groups of absorption bands in the OH-stretching region were observed in omphacites from the upper mantle and lower crust beneath the Siberian platform (Koch-Müller et al. 2004). The intensity of the peaks centered in the higher wavenumber 3600-3624 $cm^{-1}$ range were found to be strongly correlated with inclusion-rich regions within the pyroxene crystals. TEM bright- and dark-field images obtained on crystal foils cut in different parts of the examined grains using a FIB, clearly showed that these bands could be associated with nm-sized inclusions of a sheet silicate within the omphacite matrix. It is worth noting that the spectra also showed a weak band in the high wavenumber range characteristic of omphacite and caused by vibration of intrinsic hydroxyl groups. Some of the inclusions were interpreted to have formed during the uplift of the host rock as a product of interaction with fluids. The intensities of the other peaks in the FTIR patterns could be related with [4]Al or octahedral vacancies at M2, respectively. As an additional test for the assignment of the higher wavenumber bands to a phyllosilicate phase, high-pressure spectra were collected using a diamond anvil cell up to 12 GPa and the behavior of the OH bands with pressure was compared with those of chlorite and omphacite. The OH content in the samples could be quantified using polarized radiation measurements on oriented sections, and was found to vary from 31 to 514 ppm, the lowest value from the pyroxene in the highest pressure rock, a diamond-bearing eclogite xenolith from a kimberlite pipe. Up to 20 strongly polarized bands were observed (Koch-Müller et al. 2006) in the OH-spectra of olivines from the Udachnaya kimberlite pipe (Russia). Peaks at higher energy (3730-3670 cm$^{-1}$) were assigned to inclusions of serpentine, talc and the 10 Å phase. OH groups were correlated to point defects associated with either Si sites or vacant M1 sites. Polarized experiments showed that H was bonded to O1 and O2 oxygens, forming O-H vectors parallel to the *a* crystallographic axis. Combination of the FTIR data with SIMS (Secondary Ion Mass Spectrometer) analyses allowed the calibration of the integrated absorption coefficient $\varepsilon_i$ = 37,500±5,000 L mol H$_2$O cm$^{-2}$. Later Thomas et al. (2009) pointed out that this is a wave-number independent absorption coefficient, which can be used to quantify all types of OH-defects in olivine.

Koch-Müller et al. (2003) studied the solubility of hydrogen into coesite at pressure in the 4.0 – 9.0 GPa range and temperature in the 750-1300°C range by using Al and B doped SiO$_2$ as starting materials. The OH-spectra were extremely complex with several bands, assigned to different substitutional mechanisms responsible for OH incorporation into synthetic coesite. To improve the resolution for overlapping bands, low temperature spectra were collected using a Linkam T600 freezing stage (Fig. 8). The most intense and sharp peaks were attributed to the hydrogarnet substitution $^{(T2)}$Si$^{4+}$ + 4O$^{2-}$ = $^{T2}$vacancy + 4OH$^-$, consistent with the observation that more than 80% of the dissolved water is incorporated via this mechanism. Based on the positive correlation between the content of B determined by SIMS, some of the weaker bands were assigned to B-based point defects, while others were assigned to OH groups incorporated via the Si$^{4+}$ → Al$^{3+}$ + H substitution. The water solubility was quantified using the molar absorption coefficient $\varepsilon_i$ = 190,000 Å±30,000 L mol H$_2$O cm$^{-2}$ calibrated by Koch-Müller et al. (2001) and was found to increase with both temperature and pressure. Interestingly, the incorporation mechanism for H seemed to change for pressure in excess of 8.5 GPa, at 1200°C. New sharp bands appeared in the spectrum at lower wavenumbers, while those observed for P < 8.5 GPa and assigned to the hydrogarnet substitution disappeared. On the basis of single-crystal, Raman and polarized IR, these new bands were assigned to OH groups in coesite; however both their polarization and high-pressure behavior was significantly different from that of the higher-frequency components, suggesting a different substitutional mechanism for H in coesite at higher pressures. Koch-Müller et al. (2003) also reported the first FTIR spectrum of a natural OH-bearing coesite occurring as an inclusion within a diamond from Venezuela. The FTIR spectrum of the omphacitic pyroxene associated with coesite in the diamond also showed OH-bands.

Thomas et al. (2008) studied the hydroxyl solubility in synthetic Ge analogues of the high-pressure silicates ringwoodite, anhydrous phase B and superhydrous phase B. Ge-ringwoodite was found to contain up to 2200 ppm $H_2O$, the incorporation mechanism of which was found to be different from that of Si-ringwoodite. Polarized experiments on oriented slices yielded from 2400 to 5300 ppm $H_2O$ in Ge-anhydrous phase B. Single-crystal X-ray and polarized IR measurements (Fig. 9) on oriented crystal sections showed that two mechanisms could be responsible for the incorporation of OH into this phase, i.e. the hydrogarnet substitution and the creation of vacant Mg sites (Fig. 10). The findings of Thomas et al. (2008) however implied that the use of germanates as analogues for high-pressure silicates in experimental studies was probably non recommended for anhydrous phases, since their behavior with respect the water incorporation was probably different from the Si-counterparts.

Thomas et al. (2009) calibrated specific absorption coefficient for synthetic olivines, $SiO_2$ polymorphs and rutile-type $GeO_2$ combining FTIR, proton-proton scattering, Raman microspectroscopy and SIMS. One interesting conclusion of their work was that for $SiO_2$ phases the absorption coefficient is not related with the vibration frequency of the OH point defect, but with the structure type. A mean $\varepsilon_i = 89,000 \pm 15,000$ L mol $H_2O$ cm$^{-2}$ could be determined for a suite of quartz samples with varying OH defects, while much higher values were obtained for the denser polymorphs coesite and stishovite (Fig. 11). According to the work of Thomas et al. (2009) the negative correlation between the absorption coefficient and the OH band position (Paterson 1982, Libowitzky and Rossman 1997) does not hold for NAMs, thus for quantitative purposes specific coefficients must be independently calibrated. This point is particularly crucial when the estimation of the water content of high-pressure minerals is used for geophysical modeling. Thomas et al. (2009) also present a routine to quantify water contents in NAMs based on Raman spectroscopy. An improved version of the routine with respect to the correction procedure to use in case of minerals containing heavy atoms is described in Mrosko et al. (2011).

The superior lateral resolution of SR-FTIR with respect the Globar source was exploited by Feenstra et al. (2009) to examine the zoning of hydrogen in zinc-bearing staurolite from a high-P, low-T occurrence from Samos (Greece). SIMS data showed higher hydrogen concentrations in crystal cores compared to rims, the OH content being negatively correlated with Al. The same kind of zonation was obtained using SR-FTIR on slices cut out from the crystals using a FIB. Interestingly, Feenstra et al. (2009) found that the absolute H concentration derived by FTIR were systematically lower to those derived by SIMS by about 25%. They interpreted this feature as due to a water loss at the crystal surface during FIB machining.

Piccinini et al. (2006a, 2006b) examined the polarization behavior of the OH-stretching band of phlogopites by tilting the mica flake under the polarized SR beam. They observed a strong decrease of the absorbance intensity as a function of the angle between the (001) cleavage plane and the IR beam (Fig. 12).

SR-FTIR spectroscopy for the analysis of extraterrestrial materials, such as interplanetary dust particles (IDPs) was aimed at defining the global mineralogy of the samples, for example the presence of silicates *vs* carbonates or phyllosilicates. In these studies the use of synchrotron radiation was needed because of the extremely small dimension of these samples, typically 5-10 μm in size. Most SR-FTIR studies were however devoted to the capability of IR in detecting light-elements molecular arrangements such as $OH/H_2O$ or C-H groups in the specimen. Flynn et al. (2002, 2004) reported for the first time the evidence for the presence of aliphatic hydrocarbons and of a ketone group in IDPs, by detecting $CH_2$, $CH_3$ and C=O functional groups in acid-etched particles, while Matrajt et al. (2005) measured the $CH_2/CH_3$ ratio in several IDPs and compared it to the $CH_2/CH_3$ ratio in diffuse interstellar medium (DISM). Flynn et al. (2004) found that IDPs may be composed by a significant fraction (up to 90%) of organic components, leading to estimate that, in the current era interplanetary dust contributes ~15 tons/year of unpyrolized organic matter to the surface of the Earth, and that during the first 0.6 billion years of Earth's history, this contribution is likely to have been much greater.

Probably the most common use of SR-FTIR in mineral sciences has been for high-P studies with a DAC; these studies were aimed at examining behavior and stability of important rock-forming minerals at high pressure and studying phase transitions of minerals as a function of pressure. For these researches SR is particularly useful for its brilliance in the far-IR region. Koch-Müller et al. (2005) showed that superhydrous phase B, one of the candidate minerals to occur in the Earth's mantle, exists in two polymorphic forms. The OH-spectrum of lower T one, with space group *Pnn*2, consists of two well-defined bands, while the OH-spectrum of the higher T phase, with space group *Pnnm*, consists of a single broad band. The band shifts as a function of pressure in the MIR and FIR regions indicated the HT polymorph to be more compressible than the lower symmetry, LT polymorph. The thermodynamic properties of chloritoid were derived by Koch-Müller et al. (2002) from detailed band assignment in the powder IR spectra and based on the pressure dependence of bands in the MIR and FIR spectral range. Scott et al. (2007) collected high-pressure spectra of lawsonite in the NIR region up to 25 GPa to constrain the Grüneisen parameters and the vibrational density of states under pressure. Pressure-induced FIR mode shifts were consistent with phase transitions at 4 and 8.6 GPa, respectively, in accordance with previous MIR, Raman, and X-ray studies. The 8.6 GPa transition, in particular, could be clearly identified by abrupt slope changes and the appearance of a new feature in the spectrum near 368 cm$^{-1}$. High-pressure infrared spectra of talc were collected (Scott et al. 2007) up to 30 GPa in the whole 150 to 3800 cm$^{-1}$ range (Fig. 13). Significant changes in relative intensities were observed in the FIR region. The pressure shift of the hydroxyl vibration of talc was non-linear with a faster rate at higher pressures, the average pressure shift being close to 2.1 cm$^{-1}$/GPa. The ambient pressure spectrum was fully reproduced upon decompression for both minerals, implying that all pressure-induced structural changes are reversible. This being the case, both minerals could metastably transport water to depths in the Earth corresponding to pressure well beyond the known thermodynamic decomposition conditions for these phases. HP spectra presented by Liu et al. (2003) showed that the crystal structures of both OH-chondrodite and OH-clinohumite were preserved up to 38 and 29 GPa, respectively. Broadening of the absorption bands suggested however increasing disordering of the silicate framework at high pressure (Fig. 14). For both minerals, three bands are observed in the OH-stretching region at ambient conditions. For increasing pressure all bands shift linearly to higher frequency; above 18 GPa, the slopes for the three OH bands are significantly different as a result of different degrees of hydrogen bonding.

Koch-Müller et al. (2011) performed high-pressure SR-FTIR measurements of synthetic hydrous ringwoodite ($Mg_xFe_{1-x}$)SiO$_4$ with x = 0.00 to 0.61, using three different pressure - transmitting media. All samples loaded with CsI powder or liquid argon showed a sudden disappearance of the OH bands and discontinuities in the pressure shift of the lattice modes between 10 and 12 GPa. When using liquid argon annealed at 120°C and 8.6 GPa as a pressure medium for the same samples, the OH bands and the lattice vibrations could be observed linearly shifting up to 30 GPa. The disappearance of the OH bands for non-hydrostatic conditions was interpreted as due to a stress-induced proton disordering in ringwoodite (Koch-Müller et al. 2011).

Iezzi et al. (2006) examined the high-pressure behavior of synthetic amphibole Na(NaMg)Mg$_5$Si$_8$O$_{22}$(OH)$_2$. The data showed a $P2_1/m \leftrightarrow C2/m$ phase-transition at 20-22 GPa (Fig. 15). Upon release of pressure, the room-pressure pattern is immediately recovered indicating that the pressure-induced phase-transition is reversible. By analogy with structurally related pyroxenes, Iezzi et al. (2006) suggested the existence of a new $C2/m$ amphibole polymorph stable at high pressure and characterized by fully kinked double-chains. The high-pressure behavior of amphiboles in the same system, but with increasing Li for Na substitution at the B-sites was studied by Iezzi et al. (2009) who found that the pressure at which the $P2_1/m$ to $C2/m$ phase transition occurs was linearly correlated to the aggregate B-site dimension.

The pressure-dependent behavior of strontium feldspar and wadsleyite was studied up to 24 GPa by Mrosko et al. (2011). Polarized measurements yielded 1100 ppm and 12500 ppm of water, respectively for these two minerals. A new microscope was developed for high-P measurement in the THz/FIR range. A $I2/c$ to $P2_1/c$ phase transition was detected at 6.5 GPa for strontium feldspar, while a transition at 8.4 and 10.0 GPa was observed for hydrous and dry wadsleyite, respectively. These results highlighted that the presence of water in NAMs may have a significant effect on their transition behavior with pressure. More recently, Noguchi et al. (2012) studied the pressure-induced amorphization of antigorite. They observed an anomalous behavior of the bands due to the outer OH groups with respect to the bands due to the inner OH groups upon compression. Amorphization of the sample was observed for conditions in excess of 300°C and 25.6 GPa; interestingly, the spectra indicated that OH groups were still retained in the amorphized material.

Synchrotron infrared microspectroscopy has made in the recent years considerable impact in archaeology, archeometry and in the study of cultural heritage materials because of its capability as a reliable and non-destructive analytical tool. The most important results are related to the study of ancient painting materials (Salvadó et al. 2005), which are typically multi-layered and structurally heterogeneous, consisting of various fine-grained mixtures of mineral pigments and binding media. *In situ* characterization of the various components in these materials is essential to extract information on the techniques used in their manufacture, the origin of the materials used, and to design appropriate restoration procedures. A combination of X-ray diffraction/absorption, Raman spectroscopy and SR-FTIR spectroscopy has provided considerable advancements in these studies. Notable investigations involved the analysis of frescos in the Acireale (Sicily) cathedral (Barilaro et al. 2005), Romanesque wall paintings from Spain (Salvadó et al. 2008), the characterization of prehistoric polished serpentinite artefacts (Bernardini et al. 2011) and the study of the alteration of silver foils in medieval painting from Museums in Spain (Salvadó et al. 2011). [forse c'e' qualcosa fatto da Gianfelice a Diamond]

# FTIR IMAGING

## Introduction

There are essentially three possible experimental set-ups in FTIR microscopy: (1) point detector analysis in a confocal layout, (2) FTIR *mapping* done by integrating the signal from successive locations of the specimen surface, (3) FTIR *imaging*, performed with bi-dimensional arrays such as focal plane array (FPA) detectors. The main advantage of observing an entire field of view at the same time allows spatially resolved spectroscopy of large and multi-phase samples (like polycrystalline rocks), or monitoring dynamic processes in real time.

The single spot analysis can be performed both in transmission and reflection mode. The beam size typically ranges from 100 μm to 30 μm using a conventional source, and, as discussed above, can be reduced down to 3-5 μm using a synchrotron radiation source. To characterize the spatial distribution of an absorber across a sample, a standard practice is to isolate a small area of interest using apertures placed before or after (or both) the sample, and then collect several spectra along traverses. Guilhaumou et al. (1998) presented the first application of this technique using SR-FTIR microscopy to monitor the evolution of the OH peak with a small aperture, and follow the OH diffusion profile across a pyroxene single crystal experimentally treated by Ingrin et al. (1995). More recent applications of the single spot analysis have been published by Castro et al. (2008) and Feenstra et al. (2009, reviewed above). Castro et al. (2008) measured water concentration profiles around spherulites in obsidian using synchrotron radiation with a spot size of 2 μm. The distribution of OH groups surrounding the spherulites was found to reflect the expulsion of water during crystallization of an anhydrous mineral assemblage replacing the spherulite glass. The concentration profiles were found to be controlled by a balance between the growth rate of the spherulites and the diffusivity of H throughout the rhyolitic melt, thus allowing determination of the kinetics of the spherulite growth.

In the FTIR mapping mode one measures the infrared spectrum at each in-plane point and then uses peak heights, peak areas, the integer performed in a defined spectral region or other criteria to visualize the distribution of the target molecule. This experimental set-up can be coupled to a confocal-like set-up (Minsky 1988) to achieve a high statistic and a high SNR. However, the method involves collecting a large number of single spectra and an accurate motion of the sample on the stage, thus resulting in extremely long acquisition time, up to several hours, to collect data from large areas, say few mm.

In the FPA imaging mode, one obtains the whole image in a single data collection thanks to a multichannel detection similar to the concept of recording images with charge-coupled devices (CCDs) in optical microscopy. The number of pixels (the detectors) and their effective size will depend on the FPA type. Typical arrays, designed for conventional sources, range from the 64x64 channels, providing 4096 individual spectra, to 256×256 or 1024x1024 channels allowing imaging of large areas. These are however not optimized to match a synchrotron radiation source that, because of its brilliance, may illuminate only a limited portion of these bidimensional detectors (Petibois et al. 2010a,b). Such arrays are coupled with 15X or 36X objectives, so that with a single image up to few square mm, a single pixel corresponds to a physical dimension in the range 2.5 to 5 μm.

Among the most recent attempt to push forward the imaging capability using synchrotron radiation, we can cite the IRENI beamline at SRS (Madison, USA) that collects 320 hor. x 25 vert. mrad$^2$ of IR radiation from a dedicated bending magnet. The main goal of this project is to reduce the acquisition times of IR maps by using an FPA illuminated by several beams coming from multiple optical systems. The incoming radiation is separated in 12 beams and rearranged into a 3 x 4 beam bundle with the help of a total of 48 mirrors and is then sent into an IR microscope equipped with a 128x128 pixels FPA (Nasse et al. 2011). Due to the use of multiple brilliant synchrotron beams and image reconstruction methods based on Fourier-based deconvolutions, this optical system provides a very high spatial resolution at the diffraction limit for all wavelengths, a very short acquisition time and a high SNR.

H-C-O functional groups are characterized by highly polar bonds and absorb infrared radiation with a high efficiency, therefore FTIR micro-spectroscopy coupled with imaging possibilities may be used to qualitatively and quantitatively measure these molecular arrangements in geological materials (both minerals and glasses) with a high-spatial resolution (Pironon et al. 2001, Della Ventura et al. 2010). Although FTIR imaging has rapidly evolving as a tool in biological and biochemical sciences (Heraud et al. 2007, Dumas and Miller 2003, Burattini et al. 2007, Petibois et al. 2009, 2010a, 2010b), very few applications have so far been published in Earth Sciences (Della Ventura et al. 2010). We will review in the following the available data.

**The distribution of H and C in minerals**

High-pressure nominally anhydrous minerals contain trace hydrous species (see the paragraph above and reviews of Skogby 2006 and Beran and Libowitzky 2006). However, recent studies have shown that structural hydrous species, such as OH, $H_2O$ and $CO_2$ can also be incorporated in most nominally anhydrous low-pressure minerals in the crust (Johnson 2006, Della Ventura et al. 2008a, b, Bellatreccia et al. 2009 among the others). It follows that careful analysis of trace volatile species in these minerals may provide qualitative but also quantitative information on topics such as water and $CO_2$ activity, oxygen fugacity and fluid composition in the mineralizing system. In particular, the analysis of volatile traces in volcanic materials, both minerals and glasses, may provide significant constraints on the genesis and evolution of magmatic systems (De Vivo et al. 2005).

The analysis of water in minerals (Libowitzky and Rossman 1996, Aubaud et al. 2007) and glasses (Ihinger et al. 1994, Di Matteo et al. 2004, Aubaud et al. 2007) using FTIR spectrometry is now a relatively routine technique. On the contrary, the spectroscopic analysis of $CO_2$ is common in glasses (e.g. King et al. 2002, Behrens et al. 2004, Morizet et al. 2010), while rare tests have been made for fluid inclusions within minerals (Linnen et al. 2004). Work on minerals has been so far restricted on few cases: beryl and cordierite (Wood and Nassau 1967, Armbruster and Bloss 1980, Kolesov and Geiger 2000, Khomenko and Langer 2005, Della Ventura et al. 2009, 2012), feldspathoid minerals (Della Ventura et al. 2005, 2007, 2008a, Bellatreccia et al. 2009, Balassone et al. 2012), phyllosilicate minerals (Zhang et al. 2005) and some particular forms of hydrous silica (Kolesov and Geiger 2003, Viti and Gemmi, 2009).

An issue of extreme interest regarding volcanic materials is the distribution of the volatile constituent across the crystal, which can provide insight into the evolution of the crystallizing system with time; such possibility is now offered by modern FTIR imaging capabilities. Della Ventura et al. (2007) studied the carbon speciation and distribution across vishnevite, a microporous mineral belonging to the cancrinite-sodalite group of feldspathoids, with ideal formula $[Na_6(SO_4)][Na_2(H_2O)_2](Si_6Al_6O_{24})$. This mineral is structurally characterized by open channels and columns of cages extending along the crystallographic *c* axis, where a variety of extra-framework cations and anionic groups may be hosted (Bonaccorsi and Merlino 2005). Infrared spectra show that most samples, and in particular the specimens from the holotype locality at Vishnevye Mts. (Urals), contain molecular $CO_2$ as main carbon species in the structural pores, while few others were found to be $CO_3$-rich. Moreover, polarized-light measurements (Della Ventura et al. 2007) show that the linear $CO_2$ molecules are oriented perpendicular to the crystallographic *c* axis in these minerals. Several single crystals from Latium (Italy) were found to be optically zoned, from milky-white to transparent. One of these crystals, 200 x 100 x 100 μm was manually extracted from the host-rock and examined using an FTIR microscope equipped with a mapping stage and a single element MCT detector. The data were collected using a square aperture of 30x30 μm$^2$. Figure 16 shows a surprising change in the carbon speciation during the crystal growth: the side attached to the host rock (milky-white) is $CO_3$-free while being rich in $CO_2$, whereas the opposite is observed for the transparent rim of the crystal. The reason for the observed phenomenon is not clear at the present, however the extreme zonation of Figure 16 is clearly connected with a major change in the physical conditions during the mineral crystallization.

Della Ventura et al (2008b) studied a set of carefully selected, transparent leucite crystals, free from any evidence of analcime alteration from the Alban Hills volcano (Rome, Italy). The purity of the specimens was checked by SEM-EDAX microanalyses across the grains, which showed Na contents systematically < 1.0 wt%. FTIR microanalyses showed for most crystals a well-defined and rather broad absorption in the $H_2O$ stretching 3000-4000 cm$^{-1}$ region (Fig. 17a), consisting of three components centered at 3604, 3500 and 3245 cm$^{-1}$, respectively. FTIR mapping under conventional light showed some crystals to be homogeneously hydrated (Fig. 17b) with $H_2O$ contents up to 1600 ppm. Other samples showed an extremely zoned $H_2O$ distribution (Fig. 17c) consisting of a completely anhydrous core with a hydrous rim containing significant (~1200 ppm) homogeneously distributed water. High-resolution mapping (Figure 17d) showed a sharp rim/core transition suggesting a sudden change in the magmatic conditions during the crystal growth.

Bellatreccia et al (2009) and Balassone et al. (2012) studied a large set of sodalite-haüyne group minerals, from a wide variety of geological occurrences. Unpolarized spectra showed intense and multi-component absorptions in the $H_2O$ stretching (4000 – 3000 cm$^{-1}$) region and, in most cases, a very sharp and intense band at 2351 cm$^{-1}$ indicating the presence of $CO_2$ molecules in the studied samples (Della Ventura et al. 2005, 2007, 2008a, Bonelli et al. 2000, Miliani et al. 2008). According with the systematic work of Bellatreccia et al (2009), sulfatic members in the group (haüyne and nosean) are typically $CO_2$-rich, while chlorine-rich members (sodalite) are systematically $CO_2$-free. FTIR maps collected by Bellatreccia et al. (2009) and Balassone et al. (2012) (Fig. 18) showed a non-homogeneous distribution correlated to micro-fractures across the crystal; interestingly, the amount of $CO_2$ was inversely correlated to that of $H_2O$ (Fig. 18).

The distribution of H and C across cordierite was examined by Della Ventura et al. (2009, 2012). Figure 19b shows a strongly zoned distribution of the $H_2O$ signal across a sample from a granulitic enclave within the dacitic lava dome of El Hoyazo (SE Spain), which contrasts with the relatively homogeneous distribution of $CO_2$ in the same section (Fig. 19a). FPA images (Fig. 19c-d) shows that most of the intensity in the 3700-3400 $cm^{-1}$ region is associated with micro-fractures within the cordierite, where secondary alteration products are present. Noteworthy, the FPA images of Figure 19e-f show that acicular sillimanite crystals enclosed within the cordierite host are occasionally hydrated. The image of Figure 19e-f provides a direct proof that using an FPA detector, at these wavelengths (2-3 μm), the spatial resolution is few μm, i.e. close to the diffraction limit. Similar results were obtained by Della Ventura et al. (2012) on samples from different localities.

**Imaging of inclusions in minerals**

Crystals may contain tiny, generally micrometric-sized and variable shaped impurity parcels within their cavities or fractures. Parcels of liquid ± vapor ± solid are know as fluid inclusions (e.g. Roedder 1984); analogously melt inclusions are droplets of glass ± vesicles ± solid (e.g. Sobolev 1996, Frezzotti 2001). Fluid inclusion studies are largely applied (Andersen and Neumann 2001) in order to define the role of fluids on the genesis and deformational processes in mantle and crustal rocks. They also allow constraining models on the genesis of ore deposits and providing information for the geothermal and petroleum explorations and production industry (McLimans 1987, Guilhamou et al 1998, 2000).

Melt inclusions are considered samples of melt that were trapped within crystals as they grew from the magma (Sobolev 1996, Frezzotti 2001, Lowenstern 1995, Danyushevsky et al 2002). Depending on favorable conditions (absence of crystallization, alteration and re-melting processes), melt inclusions preserve the composition of the melt in thermodynamic equilibrium with their host minerals. Therefore, the content of dissolved $H_2O$-$CO_2$ in the inclusions provides information related to the entrapment pressure and to the volatile budget of the magma, with implication from magma genesis and crustal storage to magma rheology and to eruptive dynamics.

Due to their typically small size, inclusions in minerals are studied using Raman spectroscopy, a technique that allows analyzing areas as small as 0.5-1 $μm^2$. The increasing availability of synchrotron-radiation FTIR facilities, and the development of FTIR imaging techniques under both synchrotron and conventional light, now allows extending the application of infrared spectroscopy to the study of this particular type of geological samples. As said above, probably the first FTIR images of hydrocarbon-rich fluid inclusions within geological samples were those of Guilhamou et al. (1998, 2000) showing the presence of $H_2O$, $CO_2$ and aliphatic molecules in their samples. Wysoczanski and Tani (2006) examined the water content distribution in silicic volcanic glasses and related melt inclusions from Sumisu Caldera and Torishima Volcano in the Izu Bonin Arc (Japan). They found that $H_2O$ measurements done with the FPA had a precision comparable to that obtained on the same locations with single-spot analyses. Similar water contents were analyzed in both the groundmass glasses and melt inclusions within honeycombe plagioclase, indicating that the inclusions underwent water loss by degassing and/or diffusion (Wysoczanski and Tani 2006).

Detailed SR-FTIR maps were collected in the water stretching region by Seaman et al. (2006) using 10x10 to 15x15 $\mu m^2$ aperture size and displacing the sample by steps smaller than the aperture across anorthoclase megacrysts and enclosed melt parcels. Based on polarized measurements on three orthogonal sections, a water content of 126 ppm was obtained for the anorthoclase. Absence of a band at 1630 $cm^{-1}$ pointed to the presence of OH groups only in the inclusion's melt, in concentrations from 0.12 to 0.39 wt%, one order of magnitude higher than the crystal host. FTIR maps showed zones of elevated water concentration at the crystal-melt inclusion boundary (Fig. 20) suggesting that water probably diffused from the inclusion into the crystal (Seaman et al. 2006).

Mormone et al. (2011) used FPA-FTIR imaging to characterize the content and distribution of C-H-O species in melt inclusions within olivines from the most primitive rocks erupted at Procida island, Phlegrean Volcanic District (Naples, Italy). Optical and SEM microscopy showed the inclusion to be mostly glassy, with few being partly to almost totally crystallized. FPA images of a sample containing only molecular $CO_2$ revealed the inclusion core to be water- and $CO_2$-free (Fig. 21). SEM data showed that the apparently homogeneous core actually was partly crystallized (thus accounting for the absence of $H_2O$ and $CO_2$) and contained several nanometer-sized crystals (possibly clinopyroxene) and bubbles. The resolution of the FPA (~3–5 μm in the $H_2O$ stretching region) was unable to resolve the glass interspersed with the crystals, however some red pixels (Fig. 21) documented the presence of local enrichments in molecular $CO_2$, corresponding to proto-bubbles of size comparable to the FPA image resolution. The spectra showed a doubled band in the 1400-1500 $cm^{-1}$ region, typical of $CO_3$ groups dissolved in the glass (Dixon and Pan 1995), and a doubled peak at 2349-2361 $cm^{-1}$ typical of gaseous $CO_2$ (Stolper and Ahrens 1987). $H_2O$ contents in the range 0.80 to 1.72 wt% in the glassy inclusions, and up to 2.69 wt% in the partly crystallized ones, could be quantified using FPA-collected spectra. These latter inclusions also showed $CO_2$ contents up to 890 ppm. $CO_3$ contents up to 2653 and 1293 ppm were also found in the glassy and crystallized inclusion, respectively. The data allowed to recalculate entrapment pressures ranging from 350 MPa to less than 50 MPa, suggesting that the magma ascent was dominated by degassing (Mormone et al. 2011).

Figure 22 from Della Ventura et al. (2010) shows the distribution of $H_2O$ in a melt inclusion entrapped within olivine phenocrysts from a scoria erupted at Stromboli (Sicily, Italy). The inclusion is relatively large (~100 μm across) but to perform the experiment the host crystal has been doubly-polished to a thickness of ~30 μm, in such a way that the inclusion is completely exposed on both sides of the crystal slice, and the beam interact only with the parcel (Fig. 22, above) without contaminations from the host. The FTIR image (Fig. 22, below) shows a roughly homogeneous distribution of water in the glass (note how the host olivine is anhydrous) and highlights the presence of a central volume with a higher water content (shrinkage bubble), probably derived from water saturation consequent to depressurization. Exsolved volatiles can generate internal overpressure that blows up the inclusion and allows gas escaping. The outer red area that appears to be linked to the inclusion in Figure 22 can represent the water outflow from the melt inclusion.

Figure 22b displays the case of a solid inclusion within a sample of F-rich edenite, $NaCa_2(Mg,Fe)_5(Si_7Al)O_{22}F_2$. The FTIR image shows the presence of a highly hydrated phase within the host F-edenite, which, on the basis of the IR signal in the OH-stretching region, can be identified as a layer silicate intermixed within the host double-chain silicate. The presence of such exsolutions may be common in this type of amphiboles, and reflect the narrow stability field of the amphibole with respect to the layer-silicate, which has similar chemistry and a very close local structure (Na et al. 1986). The FTIR image displayed in Figure 22 reveals a very interesting feature: while the host amphibole is F-rich, the associated layer-silicate is OH-rich, suggesting a strong preference of the hydroxyl component for the layer-silicate structure with respect to the edenite structure.

**FTIR imaging of dynamic processes**

As outlined above SR may open new opportunities in many research areas and in particular in time resolved imaging applications to investigate complex phenomena in real time. Actually, FTIR spectroscopy is an extremely powerful tool in high-temperature studied where exsolution of volatile species (such as dehydration) is involved (Aines and Rossman 1984, Prasad et al. 2005, Zang et al. 2006, Bonaccorsi et al. 2007 among the others). Typically, single spot data are collected *in situ*, and the band intensity of $H_2O$/OH absorptions at each step during the heating experiment is plotted as a function of the varying temperature. In such a way, dehydration curves are obtained (e.g. Prasad et al. 2005, Zang et al. 2006,

Bonaccorsi et al. 2007). In other cases, the absorbance is plotted as a function of time; in such a case, kinetic information on the dehydration process (Tokiwai and Nakashima 2010), or H diffusion mechanisms throughout the matrix (Ingrin et al. 1995, Stalder and Skogby 2003, Castro et al. 2008) are obtained.

Prechtel and Stalder (2010) synthesized pure enstatite at 6 GPa, 1250°C under water-saturated conditions and variable silica activity. Polarized FTIR measurements showed a strong pleochroic behavior of the OH-bands. The intensity variation of the observed peaks as a function of the Si/Mg ratio in the system allowed assignment of the higher- and lower-energy band to tetrahedral (Si) and octahedral (Mg) defects, respectively. FPA images were used to characterize the evolution of the pyroxene chemistry during the crystal growth by monitoring (Fig. 23) the intensity evolution of the IR bands across the samples.

The possibility to collect *in situ* images of an area of the sample during high-temperature experiments provides a new way to monitor these complex processes. An example is given in Figure 24 from Della Ventura et al. (2010), where preliminary results obtained during dehydration experiments on leucite are displayed. Figure 24a shows the images collected along a continuous ramp where the crystal fragment was heated using at a constant rate of 5°/min using a Linkam T600 FTIR heating/freezing stage. FPA images were taken *in situ* during the experiment with a 64x64 array and a 15x objective at the nominal resolution of 16 cm$^{-1}$ adding 32 scans. With this experimental set-up each image was collected in ~ 90 seconds. Although these conditions are still insufficient to study very fast reactions, they represent a clear advantage when compared to conventional spectroscopic data collection. Images refer to a crystal fragment of dimension of ~250x250 μm$^2$ and thickness 190 μm. The IR light illuminates only an edge of the fragment and the condenser of the microscope collects the light only from an area of ~170 x 170 μm$^2$, which is less than the overall size of the crystal. In this way it was possible to enhance the contrast of the water band at the edge of the section monitoring the hydratation mechanism in real time. Selected images at constant temperature interval of ~ 100°C (corresponding to ~ 20 min heating) show that the sample dehydrates smoothly, and that at ~ 400°C it is almost anhydrous. In Figure 24b a second fragment of the same starting material has been heated abruptly (50°/min) up to 300°C. Images show that when arrived at the target temperature, the sample has lost almost half of its initial water content. At constant $T$ = 300°C, the FPA images (taken at the same conditions as in Fig. 24a) show a continuous, although relatively slow, dehydration. After 150 minutes the sample is again almost anhydrous. The results of the procedure shown in Figure 24 are still preliminary, and other tests are underway to extract a quantitative behavior from the data. In the last images of the processes shown in Figure 24 a slightly not homogeneous distribution can be recognized. Different causes may be at the origin of the observed distribution, the most probable being the occurrence of a non-perfect flat sample surface. In any case, from the discussion above it is evident that the possibility to monitor the distribution of an absorber *vs.* time across the studied material, in addition to measure it, opens new opportunities for studying and understand phenomena occurring in anisotropic materials, or performing experiments with a temperature gradient set across the sample. All these processes are ~~mainly~~ "non-equilibrium" phenomena in which it is extremely difficult to identify parameters describing the system. The investigations of these complex dynamical processes with imaging techniques is probably the most suitable approach to achieve a better understanding of many phenomena of great geophysical interest.

## CONCLUSIONS

Combination of SR-FTIR and imaging techniques provides new opportunities in high-pressure studies, different types of reactions, characterization of materials of interest in Earth Science, such as the presence and evolution with crystallization of light-element (notable H and C) species in geological materials. Moreover, the high-sensitivity of modern spectrometers, coupled with the high resolution offered by FPA detectors now allows imaging of features few μm in size, such as inclusions within minerals, traditionally accessed only by Raman spectroscopy (citiamo anche IDP's?). Such features, which are hardly accessible with other micro-analytical techniques, may help in constraining the genesis and evolution of the geological system. Finally, the possibility to collect high-resolution images in a very short time, from few tens of seconds to minutes, using modern array detectors, provides a new opportunity to study many dynamic phenomena in non-ambient conditions *in situ*. Thermal treatments, dehydration processes, and other non-equilibrium processes have been already investigated in many geological and geophysical systems, but many others are just beyond the corner in order to understand Earth and environmental science phenomena.

## ACKNOWLEDGMENTS

We like to acknowledge the several people who contributed and are still contributing to the improvement of our spectroscopic work, and in particular the technical and scientific staff of the DAΦNE-L (Frascati, Rome), Diamond (Oxford) and Bessy (Berlin) synchrotron radiation facilities. Financial support was provided by PRIN 2008 to GDV. MD Welch (London) helped in improving the clarity of the manuscript.

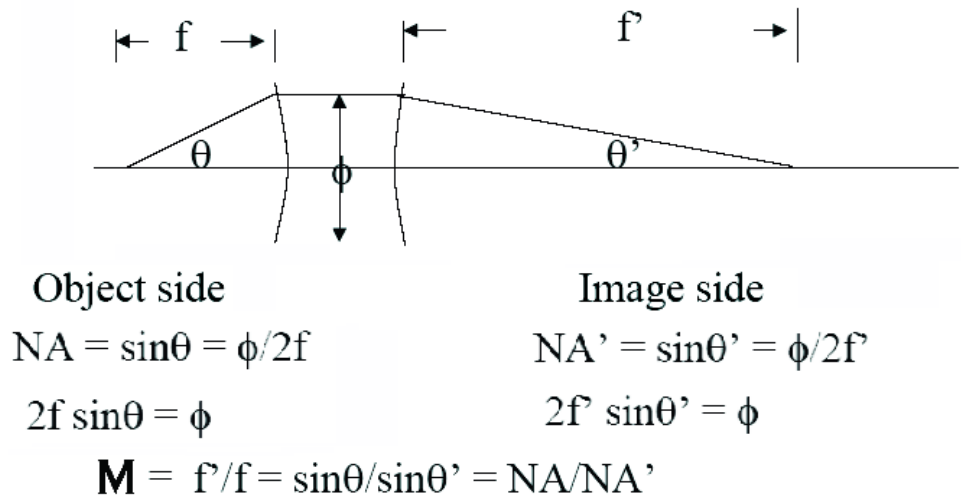

Figure 1. The magnification (M) can be defined as the ratio between the two numerical apertures (NA) of the microscope, the one on the object side (NA) and the one on the image side (NA'). Φ is the clear aperture or f-stop of the optical elements (i.e., mirrors) of the IR microscope, f and f' are the focal distances and θ and θ' their angular apertures.

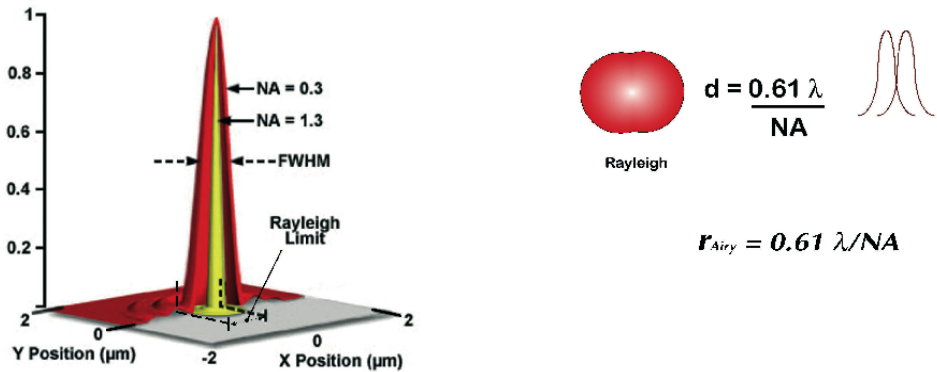
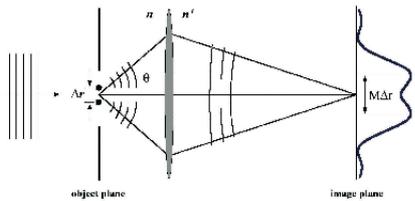
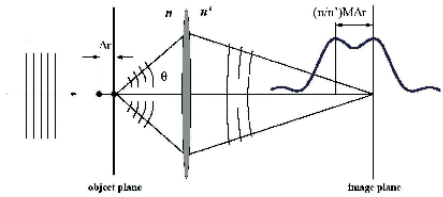
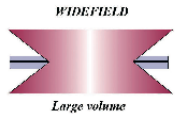
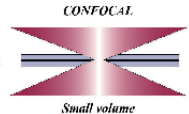

Figure 2. (top, left) Comparison of the Point Spread Function (PSF) between two optical systems with different Numerical Apertures (*NA*). According to the Rayleigh criterion (top, right), two source points can be separated when the central maximum of the Airy disc is placed at the position of the first dark fringe of the second diffraction pattern (0.61 λ/*NA*). Higher the *NA*, higher will be the spatial resolution of the system. However, in a real case, with this criterion a full separation can be obtained only if the background is negligible and the signal associated with the two source points has the same intensity, a condition that corresponds to a minimum contrast of 26.4% (e.g. Levenson et al., 2006). (Bottom) comparison of spatial lateral (left) and axial (right) resolution for confocal and widefield microscopies. Equations show that both the lateral and axial extent of the confocal PSFs are reduced by ~30% respect to a widefield illumination and that the *NA* of the objective is much more effective in the axial resolution case (~*NA*$^2$).

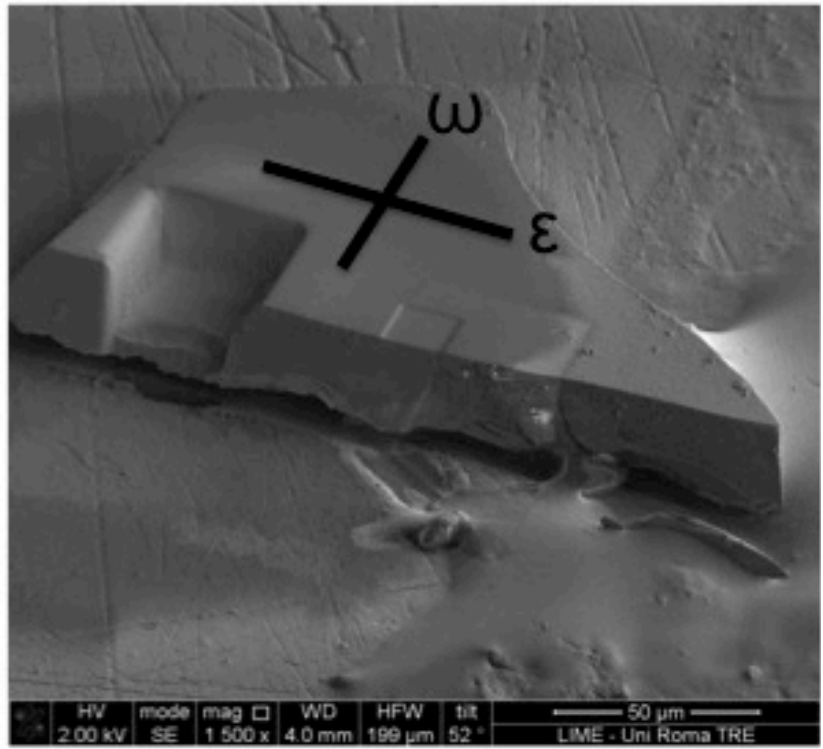

Figure 3. Oriented (hk0) doubly polished, 40 μm thick, crystal section of wardite, prepared for polarized FTIR measurements in the water stretching region (Bellatreccia and Della Ventura, unpublished). Because the transmitted IR signal is out of scale at these conditions, a hole 20x20 μm2 was machined using a FIB (Helios Nanolab, at LIME, University Roma Tre) such as at the bottom of the cavity the thickness is reduced to 8 μm. Note that the edges of the hole have been oriented such as to be parallel to the optical directions in the crystal.

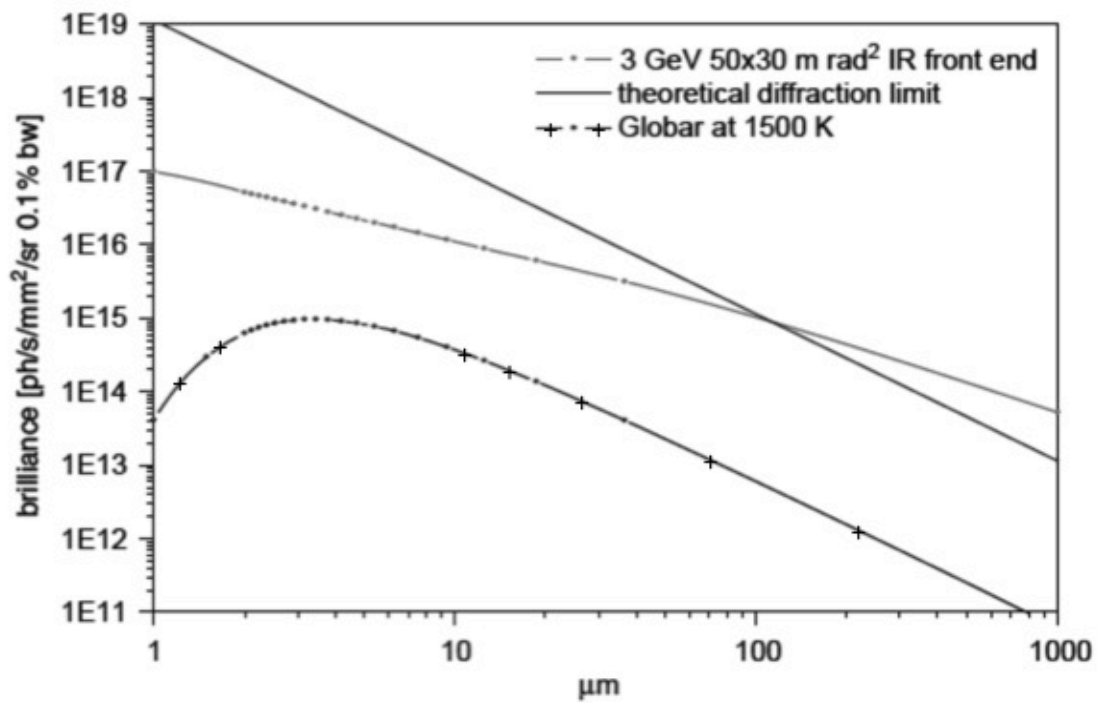

Figure 4. Calculated synchrotron radiation brilliance of a 3 GeV source in comparison to a conventional source and the ideal SR diffraction limited source (courtesy G. Cinque, from DLS, design report of the IR beamline at Diamond).

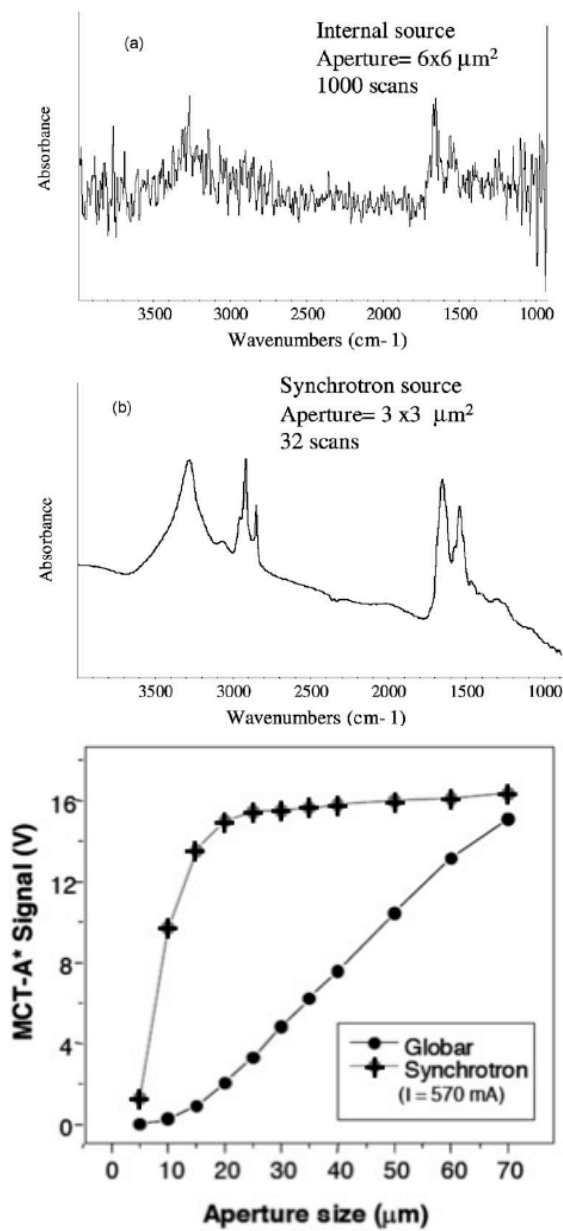

Figure 5. (top) comparison between the IR spectrum taken with a conventional source, using an aperture 6x6 μm$^2$, accumulating 1000 scans, and the SR-IR spectrum taken on the same sample location using an aperture 3x3 μm$^2$, accumulating 32 scans. (Bottom) intensity of infrared light reaching the detector as a function of the aperture size for conventional vs SR beam.

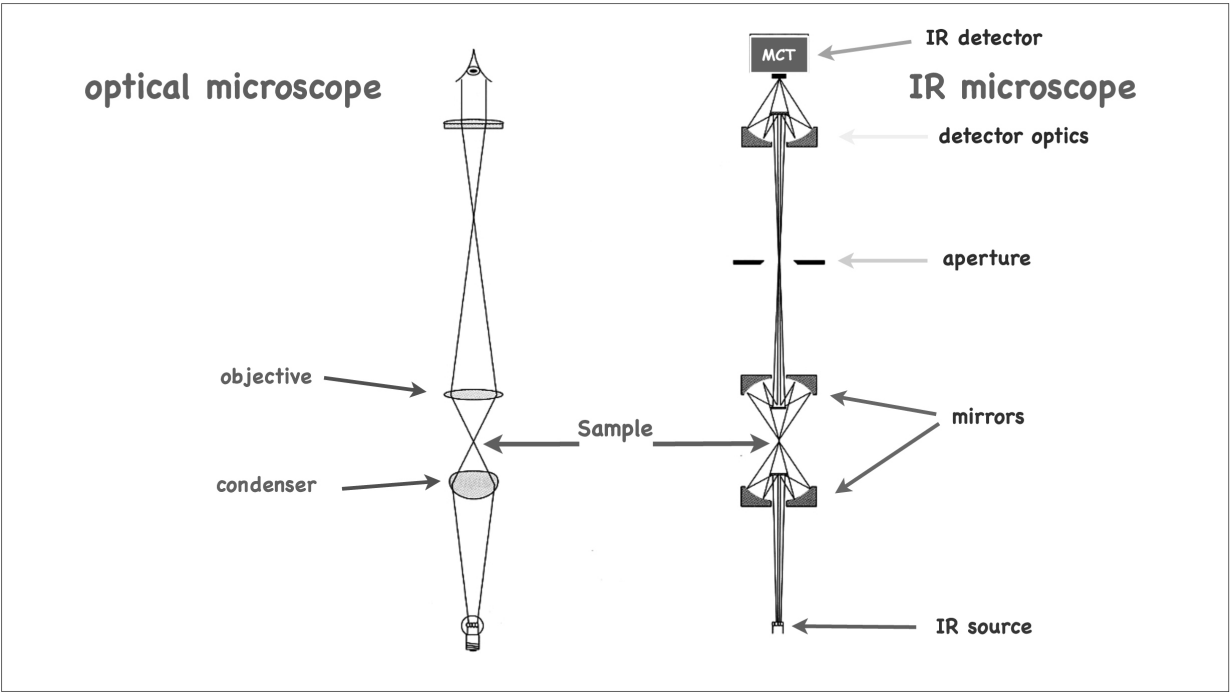

Figure 6. Schematic layout of an IR microscope compared to that of an optical microscope.

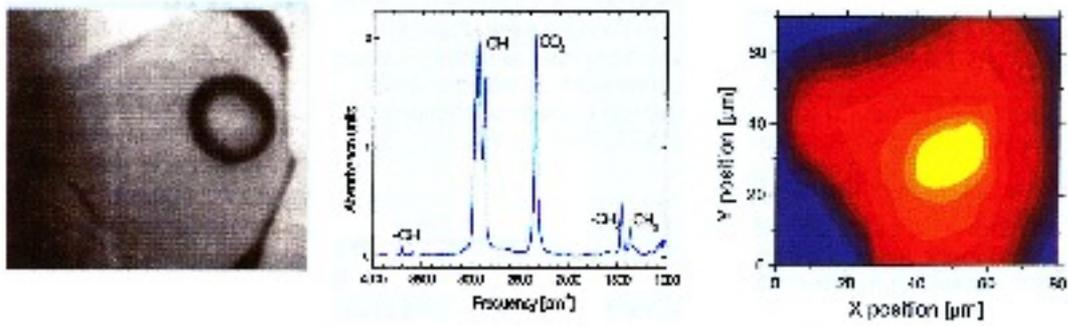

Figure 7. Infrared microspectroscopy of the $CO_2$ distribution within a fluid inclusion. Left: optical image, center: selected FTIR spectrum, right: resulting $CO_2$ map. Modified from Bantignies et al. (1998).

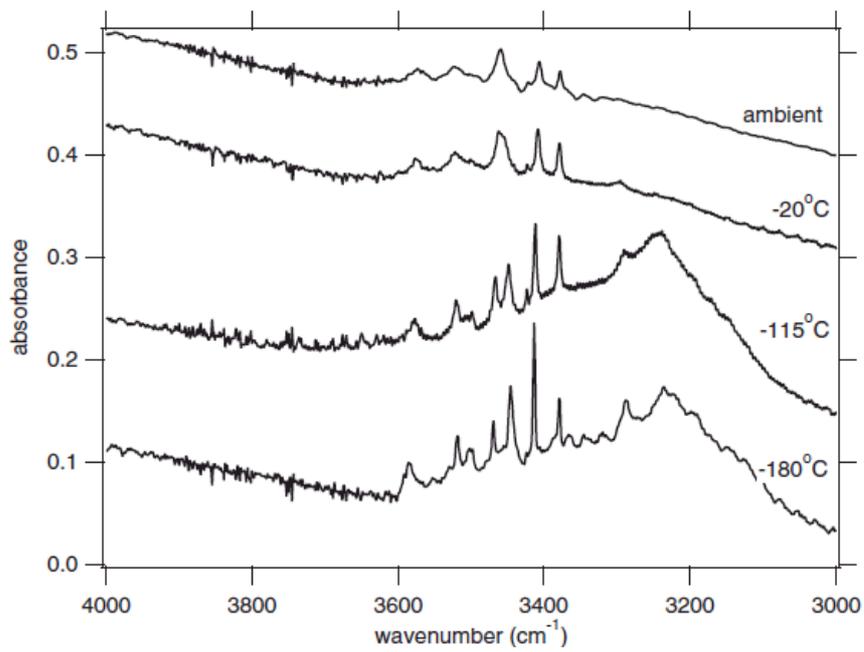

Figure 8. Unpolarized IR spectra of coesite collected at different temperatures. The broad band centred near 3200 cm$^{-1}$ is from ice condensation in the stage. From Koch-Müller et al. (2003)

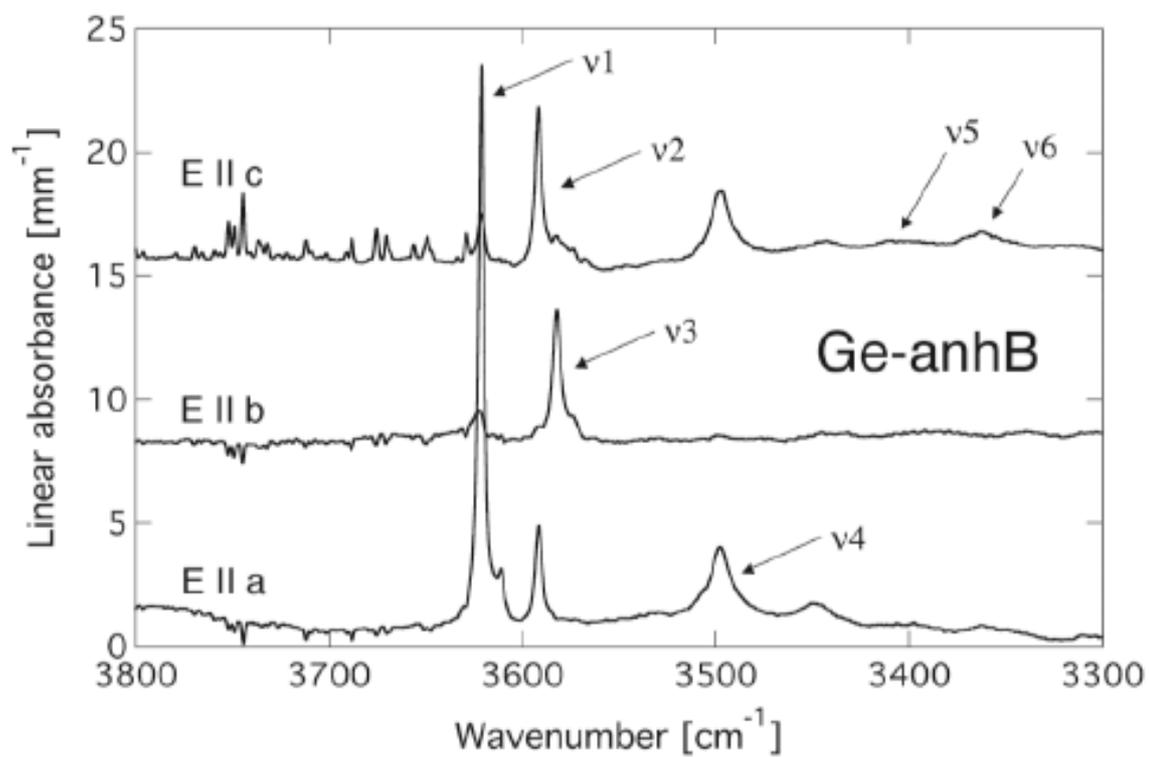

Figure 9. Polarized SR-FTIR spectra of Ge anhydrous phase B showing the strong pleochroism of the OH bands. From Thomas et al. (2008).

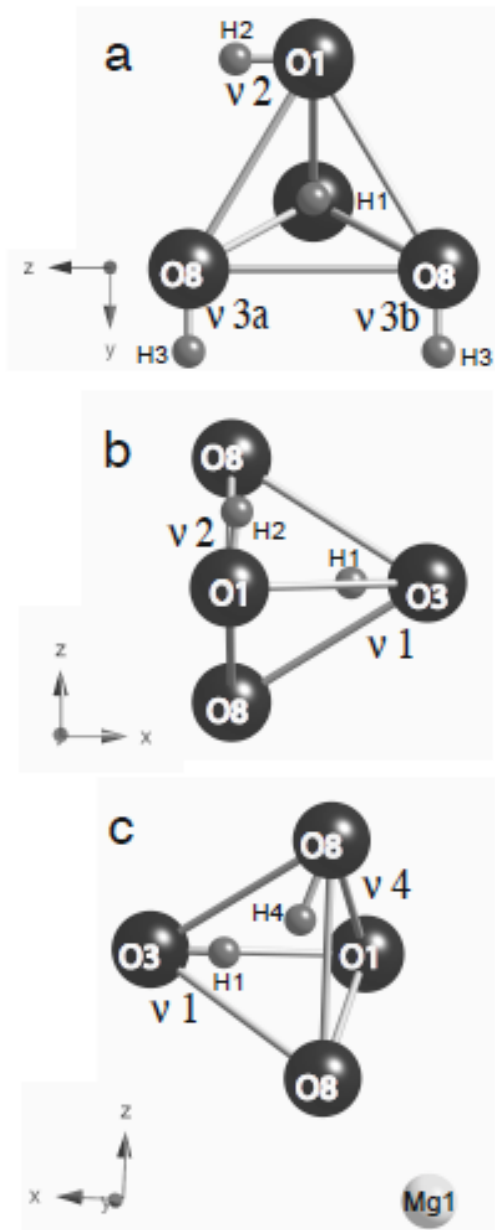

Figure 10. Orientation of the different OH groups in Ge-anhB as modeled on the basis of the polarized FTIR spectra. From Thomas et al. (2008).

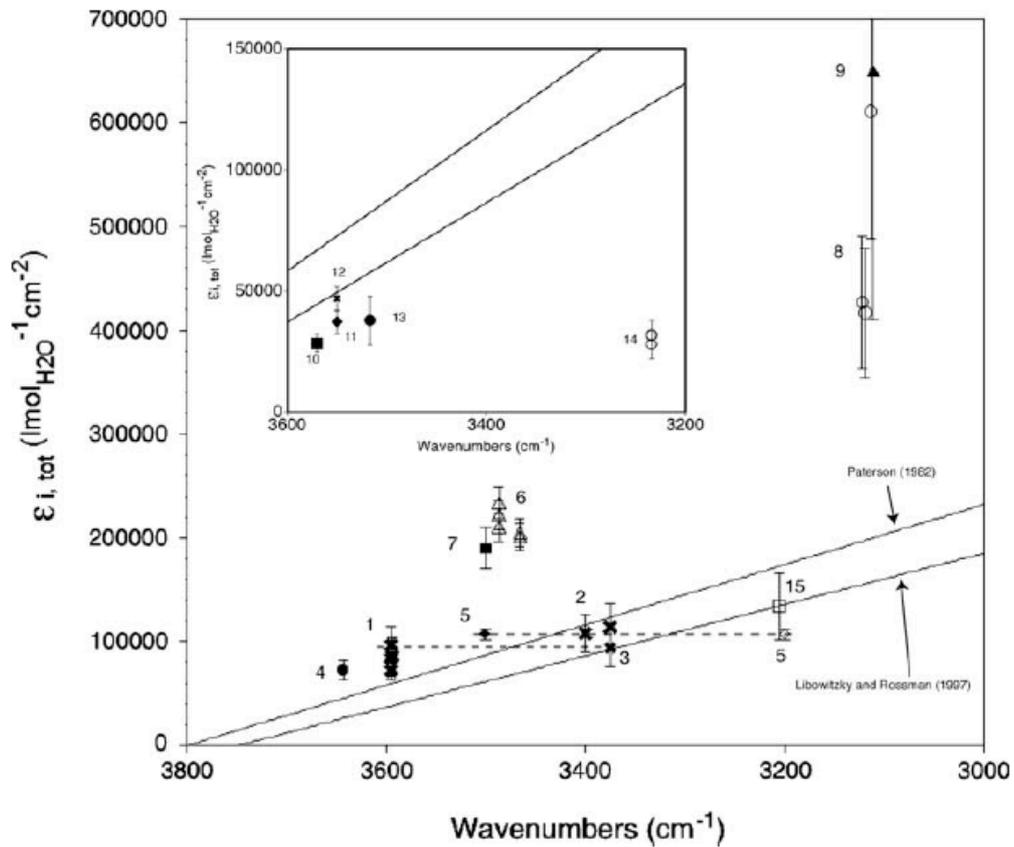

Figure 11. Absorption coefficients of doped quartz (1, 2), natural quartz (3), quartz glass (4), feldspar (5), coesite (6, 7), stishovite (8, 9) and r-$GeO_2$ (15) plotted onto the Paterson (1982) and Libowitzky and Rossman (1997) diagrams. The inset gives a magnified section of the 3,600–3,500 $cm^{-1}$ region with different olivine data. [Used with kind permission of Springer Science and Business Media, from Thomas et al. (2009), *Phys Chem Miner*, Vol. 36, Fig. 11, p. 502].

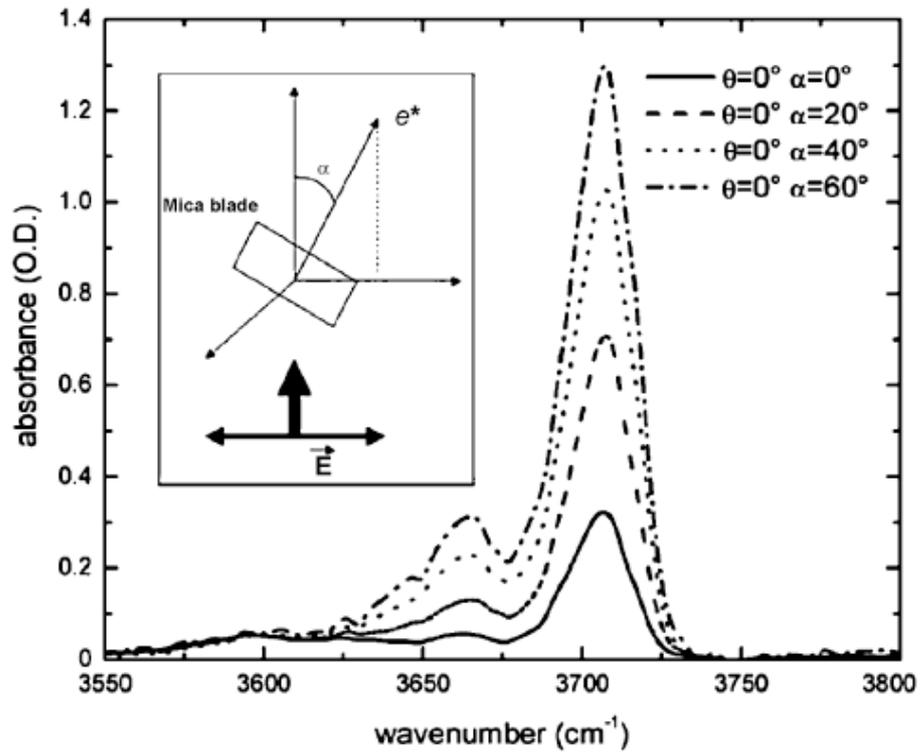

Figure 12. Polarized-light SR-IR spectra of natural phlogopite in the OH-stretching region collected by rotating the sample with respect the electric vector (see inset). [Used by permission of Elsevier, from Piccinini et al. (2006b), *Vib Spectr*, Vol. 42, Fig. 3, p. 60].

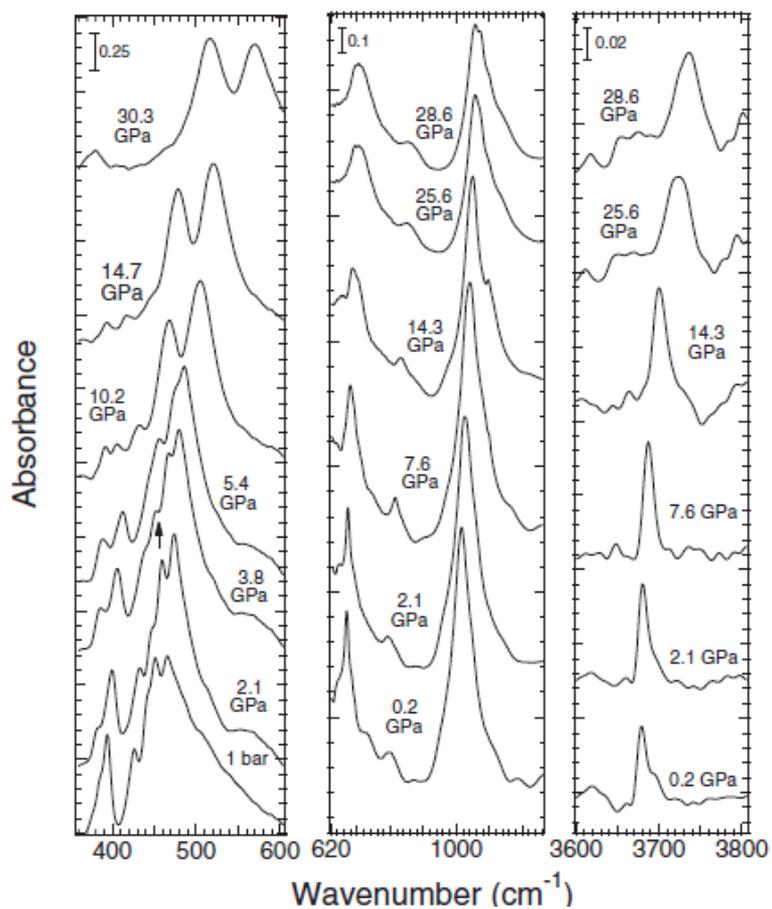

Figure 13. High-pressure SR-IR spectra of talc. Note that the ambient P spectrum is fully reproduced upon decompression. After Scott et al. (2007)

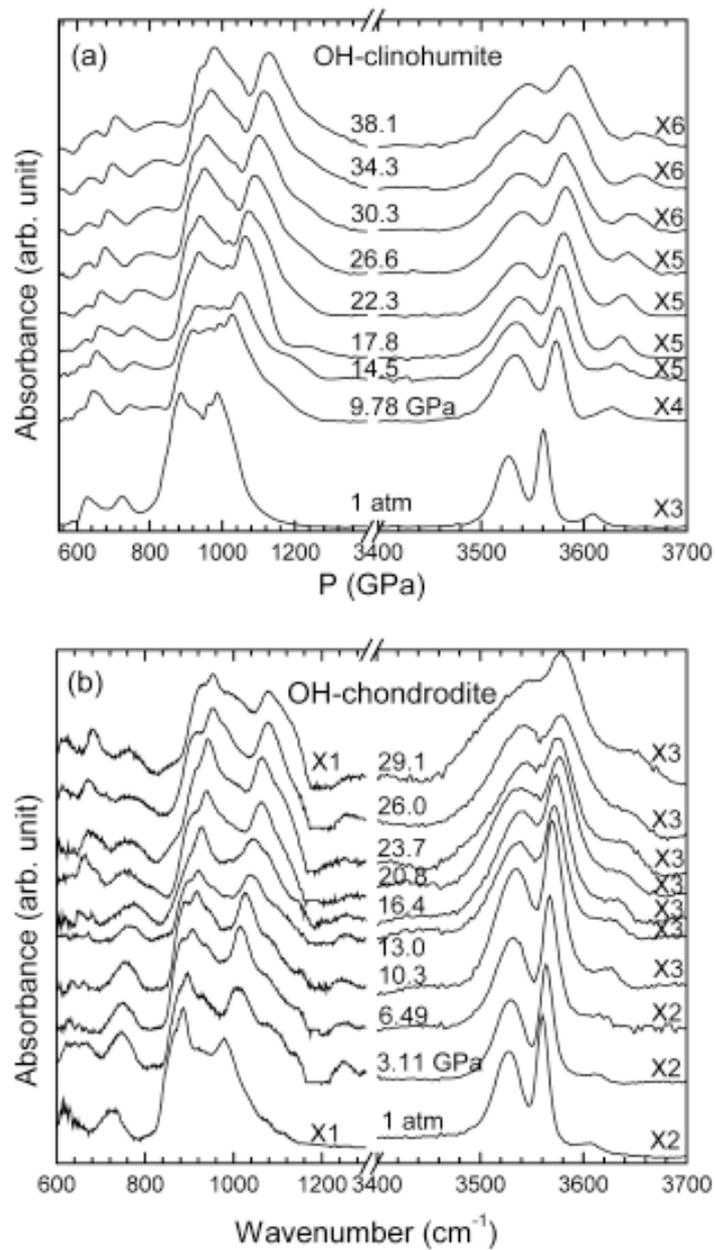

Figure 14. SR-FTIR OH-spectra of clinohumite (a) and chondrodite (b) at increasing pressure. After Liu et al. (2003).

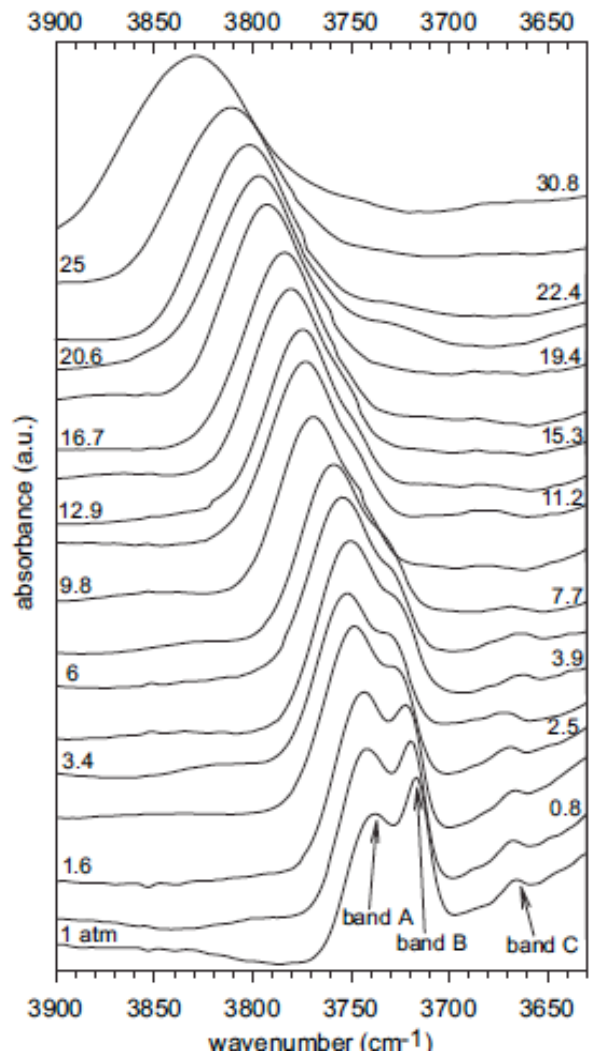

Figure 15. SR-IR high-pressure OH-spectra of synthetic amphibole Na(NaMg)Mg$_5$Si$_8$O$_{22}$(OH)$_2$. After Iezzi et al. (2006)

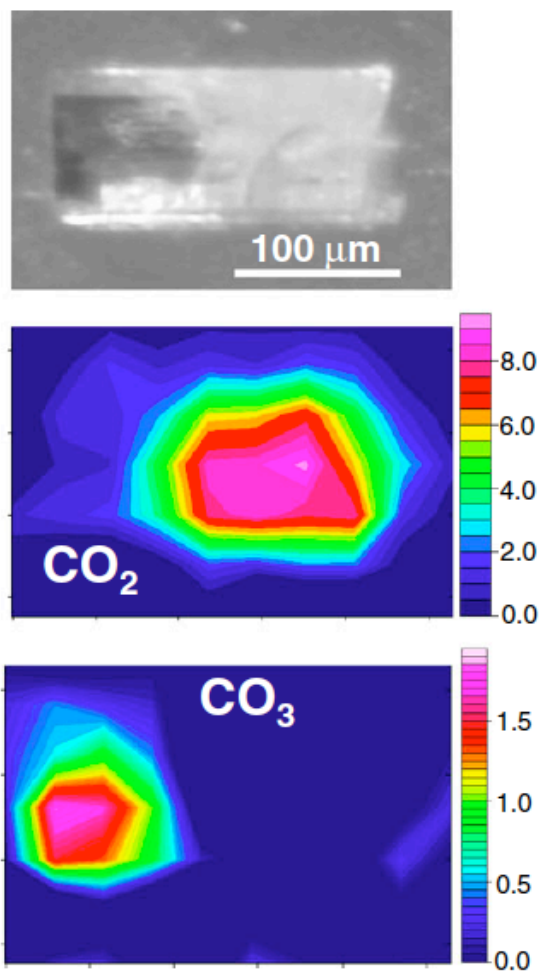

Figure 16. Top: optical micrograph of the vishnevite crystal studied. Note that the specimen is not a polished section, but a prismatic euhedral hexagonal single crystal manually extracted from the rock; the terminal pinacoid of the crystal is on the left side of the picture. Middle and bottom: FTIR mappings of the $CO_2$ and the $CO_3$ distribution across the specimen. The color intensity from blue to red is proportional to the content of the target molecule. [Used by permission of Elsevier, from Della Ventura et al. (2010), *Anal Bioanal Chem*, Vol. 397, Fig. 1, p. 2043].

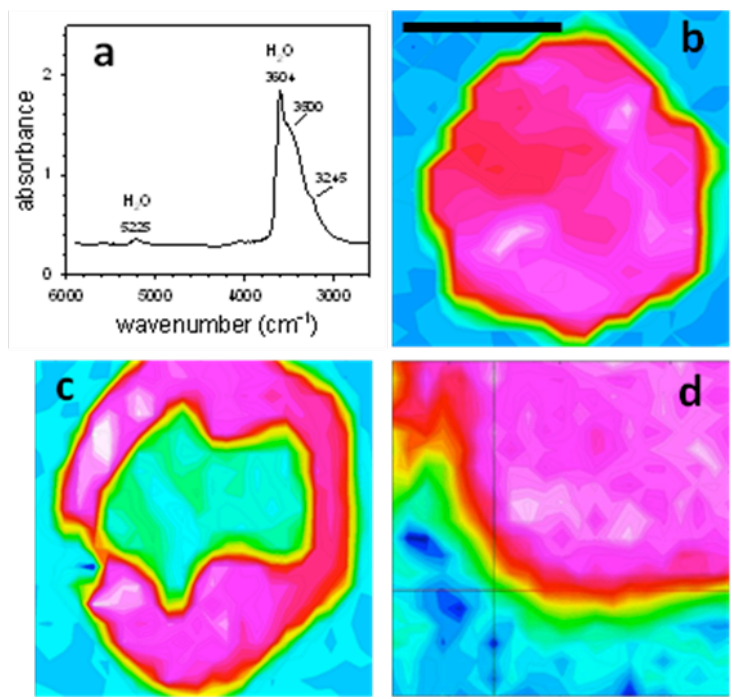

Figure 17. (a) FTIR spectrum of leucite in the NIR range. (b) homogeneous distribution of $H_2O$ across sample L131 from Tre Fontane (Rome). (c) zoning of $H_2O$ in sample LCQM2 from Quarto Miglio (Rome) showing a well-defined anhydrous core surrounded by an homogeneously hydrated rim. (d) high resolution mapping across the core-to-rim boundary for the same sample as in (c). The scale bar in (b) is 2 mm. The color intensity from blue to red is proportional to the content of the target molecule. [Used by permission of Elsevier, from Della Ventura et al. (2010), *Anal Bioanal Chem*, Vol. 397, Fig. 2, p. 2044].

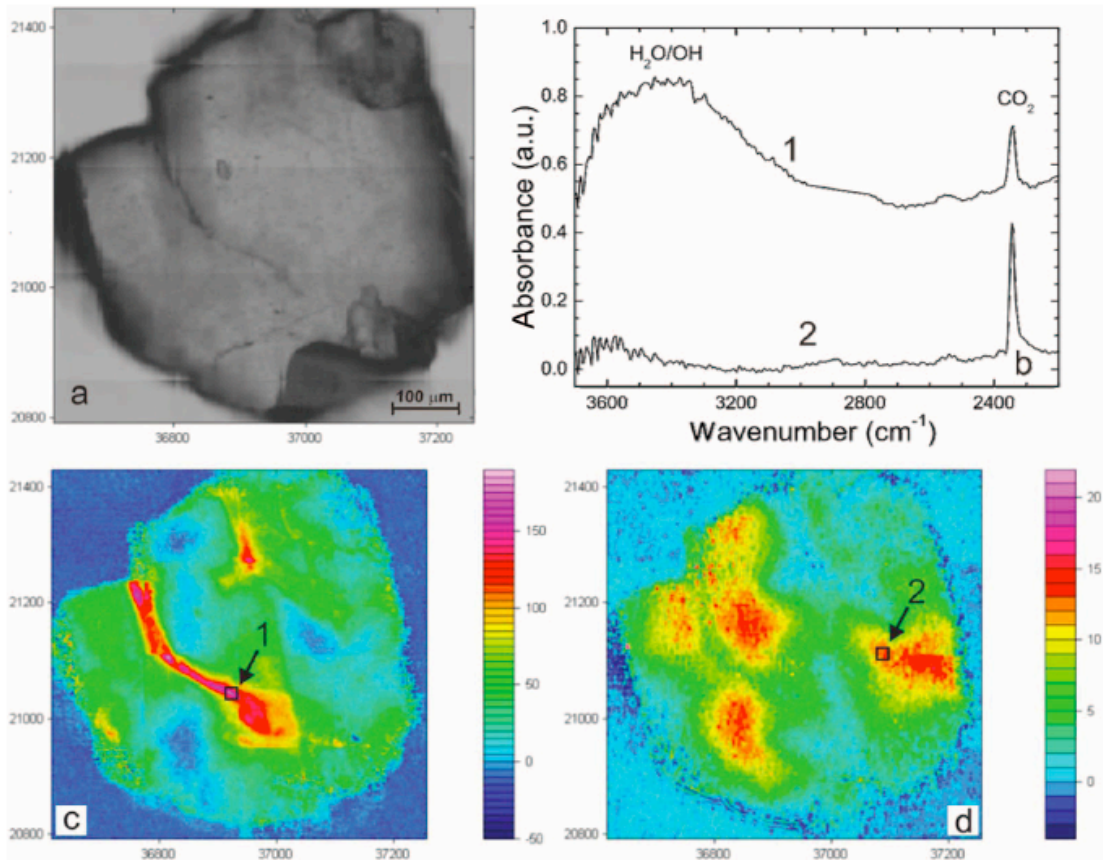

Figure 18. Distribution of H$_2$O and CO$_2$ in haüyine from Somma-Vesuvius; (a) optical image of the examined crystal, (b) selected FTIR spectra at points 1 and 2 in the (c) H$_2$O and (d) CO$_2$ images. Note that images in (c) and (d) are the result of 4 x 4 matrix of 170x170 μm FPA images. [Used by permission of Mineralogical Society, from Balassone et al. (2012), *Min Mag*, Vol. 76, Fig. 9, p. 206].

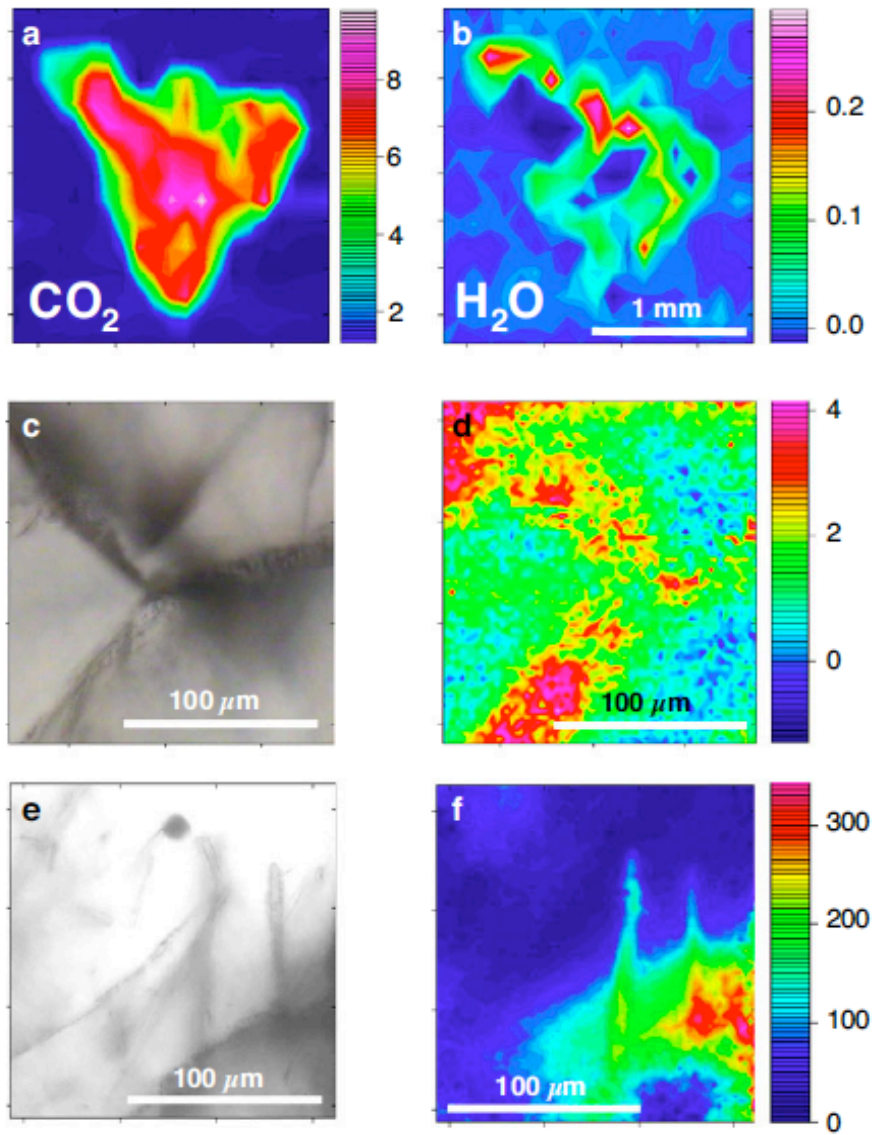

Figure 19 (a) and (b) FTIR mapping of $CO_2$ and $H_2O$ on a (010) oriented crystal section of cordierite from El Hoyazo, Spain, of thickness 140 μm. (c) to (f) FTIR FPA images of the $H_2O$ distribution in the same section; the optical image is on the left side, the corresponding FTIR image is on the right side. The image in (d) shows how the water content of this sample is associated with fractures and is probably related to alteration products, whereas the image in (f) shows that water is also associated with sillimanite needles included in the host cordierite matrix. [Used by permission of Elsevier, from Della Ventura et al. (2010), *Anal Bioanal Chem*, Vol. 397, Fig. 4, p. 2046].

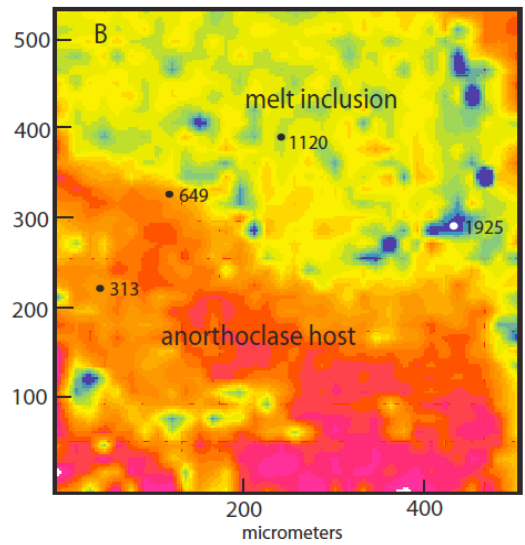

Figure 20. FTIR map in the water region across a melt inclusion/host anorthoclase crystal boundary. There is an approximately 50 μm wide zone of gradational water concentrations on the edge of the anorthoclase crystal, likely due to diffusion of water into the crystal from the melt inclusion; water concentrations shown in parts per million. After Seaman et al. (2006).

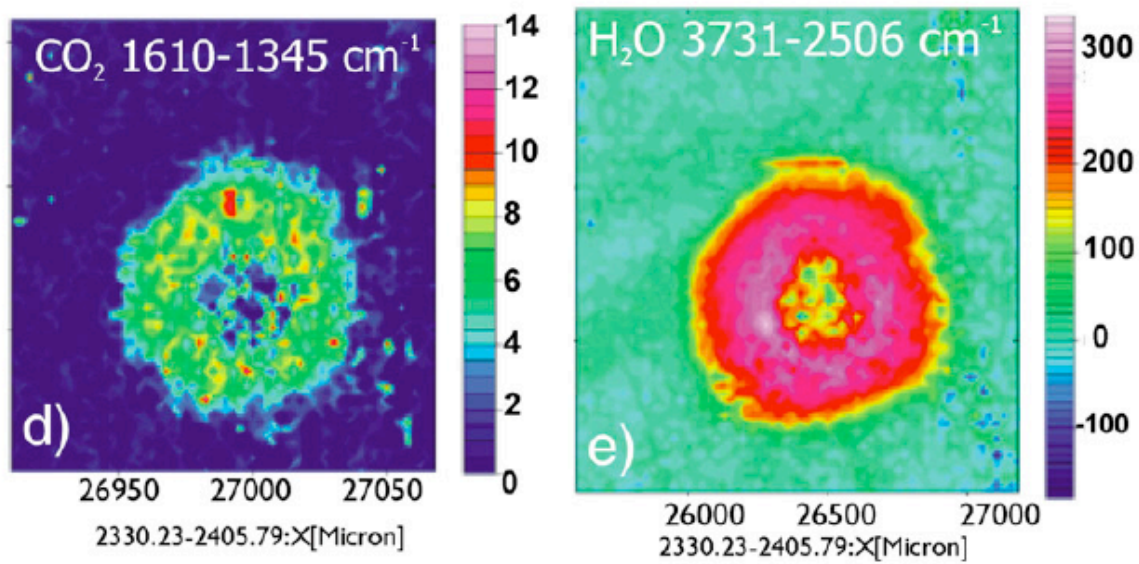

Figure 21. $CO_2$ (left) and $H_2O$ (right) distribution (FTIR-FPA images) across a partially crystallized melt inclusion within a olivine host from the Phlegrean volcanic district (southern Italy). [Used by permission of Elsevier, modified from Mormone et al. (2011), *Chem Geol*, Vol. 287, Fig. 6, p. 74].

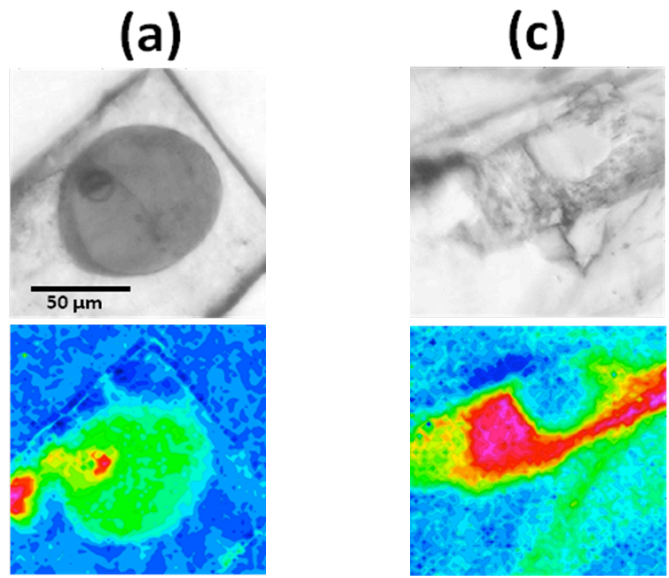

Figure 22. FTIR-FPA images (a) a melt inclusion within olivine from Stromboli (Italy) and (c) a solid inclusion within edenite from Franklin Furnace (USA). Above: optical images; below: FTIR FPA images. The scale bar is 50 μm for all images. [Used by permission of Elsevier, modified from Della Ventura et al. (2010), *Anal Bioanal Chem*, Vol. 397, Fig. 5, p. 2047].

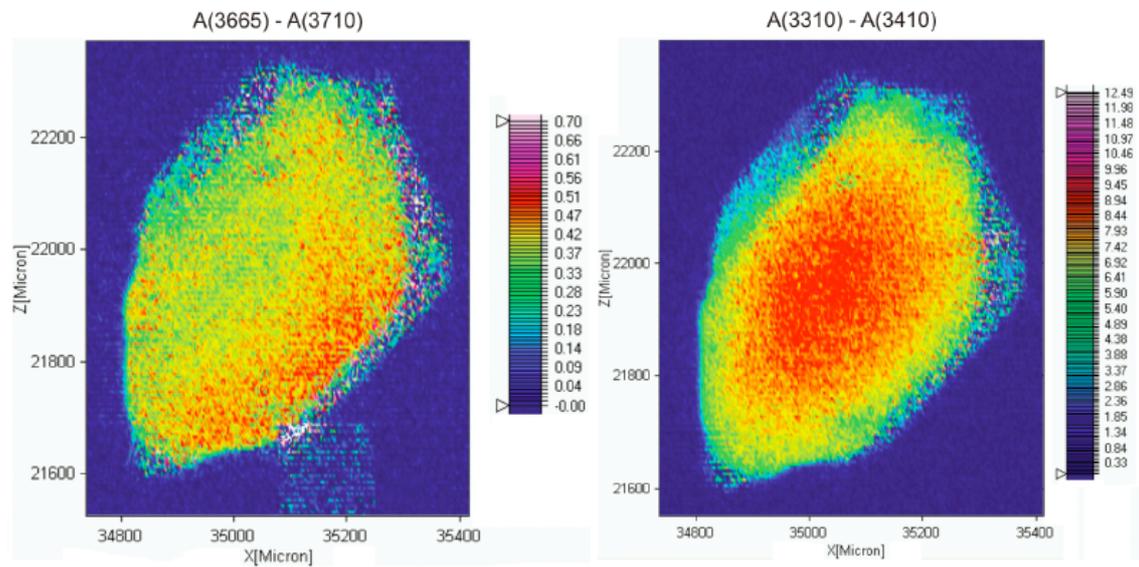

Figure 23. FTIR-FPA images of experimentally treated enstatite. OH-defects giving rise to the absorption band around 3687 cm$^{-1}$ show increasing concentration toward the rim (left), and OH-defects giving rise to the absorption band around 3362 cm$^{-1}$ show decreasing concentration toward the rim (right). Modified after Prechtel and Stalder (2010).

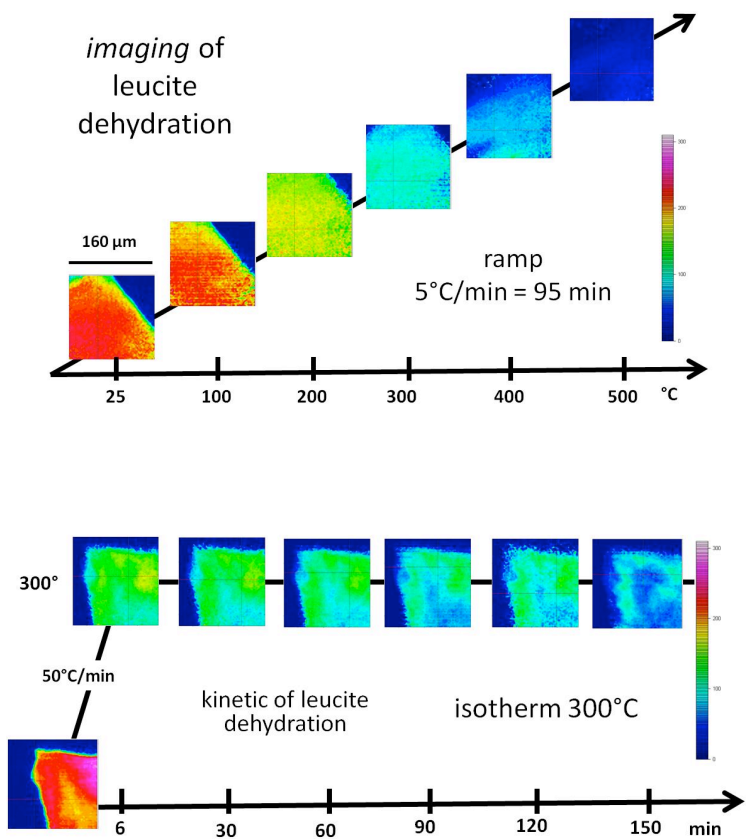

Figure 24. FTIR images obtained with a FPA detector during the *in situ* heating of a single crystal of leucite. Above: continuous ramp of 5°C/min, showing that leucite dehydrates continuously to become almost anhydrous after 400 min. Below: kinetic study of the dehydration behavior of leucite. T was increased up to 300°C at 50°C/min and than fixed at T = 300°C. The figure shows that at 300°C after about 6 min the sample lost half of its initial water content, and then continues to dehydrate. After 150 min the sample is almost anhydrous. [Used by permission of Elsevier, from Della Ventura et al. (2010), *Anal Bioanal Chem*, Vol. 397, Fig. 6, p. 2048].